\documentstyle[aps,preprint,pre]{revtex}

\begin{document}
\draft


\title{\bf Elementary electronic
excitations in one-dimensional continuum and lattice systems}

\author{D. W. Wang and S. Das Sarma} 

\address{Department of Physics, University of Maryland, 
College Park, Maryland 20742-4111}

\date{\today}
\maketitle
\pagenumbering{arabic}

\vspace*{1.5cm}

\begin{abstract}
We systematically investigate the mode dispersion and spectral weight of the
elementary excitation spectra in one-dimensional continuum and lattice 
electron systems by using the RPA, 
the Luttinger liquid model, and the Hubbard model. 
Both charge and spin excitations are studied in details
and compared among the theoretical  models. For the lattice 
Hubbard model we use both
Bethe-ansatz equations and Lanczos-Gagliano method to calculate dispersion
and spectral weight separately.
We discuss the theoretically calculated elementary excitation 
spectra in terms of the experimental inelastic light (Raman) 
scattering spectroscopy of 1D semiconductor quantum wire systems.
Our results show that in the
polarized (i.e. non-spin-flip) Raman scattering spectroscopy, only 
the 1D charge density excitations should show up with observable 
spectral weight with the single particle excitations (in RPA)
or singlet spin excitations (in the Luttinger model and the Hubbard model)
having negligible spectral weight. The depolarized (spin-flip) Raman
scattering spectra manifest the spin density or the triplet spin excitations.
We also provide a qualitative comparison between the continuum and the lattice 
1D elementary excitation spectra.
\end{abstract}

\pacs{PACS numbers: 71.10.Fd, 71.15.-m, 78.30.Fs, 73.20.Dx, 78.35.+c, 78.40.Me
}

\vfill\eject
\vskip 1pc
\section{Introduction}
There has been considerable recent interest in the elementary
excitation spectra of one-dimensional (1D) systems,
both in continuum systems
such as 1D semiconductor quantum wires (QWR) or carbon nanotubes
and in lattice models such as 1D organic conductors.
In some sense the 1D elementary excitation spectrum is remarkably simple
\cite{Haldane81,Review,dyn_response} because 
single-particle (electron-hole like) excitations 
are completely suppressed in one dimension, and only 1D collective 
(charge density and spin density) excitations exist at long wavelengths. 
In practice, however, resonant inelastic light (Raman) scattering 
(RRS) based spectroscopic studies~\cite{QWR_ref,other,2D_exciton_RRS} 
of 1D semiconductor (GaAs) quantum wires find several electronic 
excitation modes in polarized and depolarized Raman spectra whose 
mode dispersions and spectral weights have been measured carefully.
In the polarized geometry incident and scattered photons have the same
polarization vectors indicating the absence of spin flips in the electronic
excitations, while in the depolarized geometry the incident photons change
the polarization after scattering due to the spin flips of electrons.
From the many-body linear response theory, the polarized spectrum is 
proportional to the imaginary part of the charge density correlation function
and the depolarized spectrum is proportional to the imaginary part of the 
spin density correlation function. 
One particularly noteworthy feature of these Raman scattering studies 
has been the finding that the qualitative features of 1D GaAs QWR 
elementary excitation spectra (in both polarized and depolarized 
geometries) are remarkably similar to those in 2- and 3-D GaAs systems.

The goal of this paper is to investigate theoretically the charge 
and spin elementary excitation spectra as well as their spectral 
weights in 1D electron systems, both for a continuum jellium 
electron gas and for an atomic lattice model. Our calculations 
(in particular, our spectral weight calculations) should apply 
directly to the experimental Raman scattering data if 
the resonance effects are dynamically unimportant in the 
interpretation of the Raman experiments.
We refer to our calculations as the nonresonant Raman scattering (NRS)
theory where only the conduction band electrons are taken into 
account as opposed to the resonant Raman scattering where both the 
conduction band and the valence band participate. In fact, the theory 
developed in this paper has been the standard theory for discussing 
the resonant Raman scattering spectroscopy until very recently 
when several publications~\cite{our_rrs_rpa,sassetti,our_rrs_ll} 
dealing with the full subtleties of the resonance effect have 
appeared in the literature. We emphasize that, quite apart from 
the resonant Raman scattering spectroscopy, theoretical results 
presented in this paper stand on its own as a comprehensive 
theory for the elementary excitation spectra of 1D electron systems.

In this paper, we will study the standard (nonresonant) Raman
scattering spectroscopy in three theoretical models: the 
RPA-Fermi liquid model, the Luttinger liquid (LL) model, and the 1D
Hubbard model. As emphasized above, by "nonresonant"
we mean that the theory neglects all effects of the valence band in resonant
Raman scattering (which is a 2-step process, with the incident 
photon being absorbed by a valence band electron which thereby 
gets excited into an excited conduction band state with an electron 
from inside the conduction band Fermi surface subsequently 
combining with the valence 
band hole with the emission of the scattered photon).
If the valence band can be ignored, then  
only conduction band density fluctuations are 
responsible in the linear response theory of the scattering process. 
The calculation is then
simplified to the evaluation of the density-density correlation function (for
polarized spectrum) and the spin-density correlation function (for depolarized
spectrum), whose imaginary parts are proportional to the
spectra measured in the experiments.
This approach of identifying the measured elementary excitation 
spectra in the Raman scattering experiments as the charge 
(polarized spectra) or the spin (depolarized) density correlation 
function of the electron system in the conduction band has 
a long and fairly successful history~\cite{10} in the semiconductor 
structures. We take the same approach here, and construct our 
charge and spin density correlation functions (which give the 
spectral strengths of the elementary excitations through their 
imaginary parts or the corresponding dynamical structure factors) 
entirely from the conduction band carriers, ignoring all effects 
of the valence band in the resonance process.

There has been one persistent feature in the experimental Raman 
spectra of semiconductor systems, including 1D QWR structures, 
which does not seem to have an obvious explanation in terms of the 
non-resonant theory discussed in this paper. There is often a low 
energy spectral peak in the polarized spectra at an energy well 
below the expected collective charge density excitation (CDE) peak (and 
in addition to the charge density excitation peak, which always 
shows up at the usual energy). This additional peak occurs around 
the single particle excitation energy, which typically contributes 
little to the dynamical structure factor (i.e. the density correlation 
function) at the low wavevectors ($q\ll k_F$) of Raman scattering 
experiment, and therefore should have negligible (unobservable) 
spectral weight. There have been many suggestions for the resolution 
of this puzzle (namely, why the single particle excitation weight 
is enhanced in the density correlation spectrum), and we will 
quantitatively consider several of these suggestions in this paper. 
Our conclusion, based on the results presented in this paper, is 
that this puzzle in all likelihood arises necessarily from the 
resonant nature of Raman scattering experiments, as has recently 
been argued in the literature~\cite{our_rrs_rpa,sassetti,our_rrs_ll}, 
which is beyond the scope of this paper. Our critical quantitative 
consideration of the several suggested scenarios (within the 
non-resonant theory of using conduction band properties only) for 
explaining why the single particle excitation has large spectral 
weight shows that none of them is capable of resolving this problem 
quantitatively. While studying the 1D elementary excitation 
spectra is the primary goal of this paper, considering the single 
particle excitation spectral weight issue in the non-resonant 
Raman scattering is one of our important secondary goals.

We first present the results of the Fermi-liquid 
random phase approximation (RPA)
calculation for this problem in Sec. II.
The RPA calculation has been shown to give a good description~\cite{11}
for the dispersion relations of the elementary excitations in 
comparison with
the experimental results~\cite{QWR_ref}, for both
intersubband and intrasubband 1D excitations~\cite{rpa_multi}.
Being a standard Fermi liquid (FL) theory,
however, the RPA calculation is unable
to explain the relatively large spectral weight of the ``single particle
excitation''(SPE) in the polarized spectrum of the experiment as 
discussed above.
We include the effects of the breakdown of momentum conservation and
the nonparabolic energy dispersion in our calculation in Sec. II-D to 
check if they can explain the SPE feature within RPA, but neither
gives qualitatively correct results for the polarized spectrum. 
In the Luttinger liquid model we present in Sec. III, we find 
\textit{zero}
weight at the SPE energy (as we should, since in the LL model 
there are no single particle like 
quasi-particle excitations), and all the spectral strength is at the 
charge boson mode,
which is exactly the CDE mode in RPA.
In Sec. IV, we use the 1D (lattice) Hubbard model with repulsive on-site 
spin-dependent interaction to study this problem. The study of 1D 
elementary excitations and the associated spectral weights in the Hubbard 
model is one of our main results in this paper. This Hubbard model study
was originally 
motivated by the suggestion in ref.~\cite{schulz93} 
that a possible way to interpret this SPE puzzle (the existence of a
single particle peak in Raman scattering) in 1D
should be different from those in higher dimensions, and the so-called 
SPE peak may be arising
from the spin-singlet excitations (SSE) of interacting 1D systems
\cite{schulz93,schulzemery}. 
Therefore, we choose to study in detail the 1D Hubbard model, in which
the spin-dependent interaction is expected to enhance the contribution
of the spin-singlet excitation, which is proposed \cite{schulz93,schulzemery}
as the extra SPE-like feature showing up in the experiment. 
Although the Hubbard model is a lattice model
and consequently may not apply directly to the continuum QWR system,
we argue that it is useful to understand the detailed excitation
spectra in the 1D Hubbard model in the context of this problem 
because one can quantitatively study the interacting 1D elementary 
excitation spectra using the Hubbard model.

Our conclusion based on our work (as presented in Sec. II-IV) is that 
all of these NRS theories fail in providing
an explanation for the large ``SPE'' spectral weight in the long wavelength
region observed in
the experimental polarized RRS spectrum,
although there could be substantial spectral weight in the single particle
excitation at $large$ wavevectors (short wavelengths).
The spectra of depolarized nonresonant Raman
scattering is also studied in these different models, and good general
agreement with experiment is found in the sense that the depolarized 
spectra describe the spin density excitation (SDE)
or the triplet spin excitation. 
\section{Fermi liquid model}
In the standard many-body theory, which neglects all interband resonance
effects and considers only the conduction band, the NRS intensity for polarized
spectrum is proportional to the imaginary
part of the charge density correlation function $\chi_\rho$: 
\begin{equation}
\frac{d^2\sigma}{d\Omega d\omega}\propto \frac{1}{\pi}\mathrm{Im}
\chi_\rho(\it{q},\omega),
\end{equation}
where
\begin{equation}
\chi_\rho(q,\omega)=i\int_0^\infty dt e^{i\omega t}\chi_\rho(q,t),
\end{equation}
and
\begin{equation}
\chi_\rho(q,t)=\langle\left[\rho^\dagger(q,t),\rho(q,0)\right]
\rangle_0,
\end{equation}
where $\rho(q,t)\equiv n(q,t)-n_0$ is the density fluctuation operator and
$n(q,t)\equiv\sum_{ps}c^\dagger_{p+q,s}(t)c_{p,s}(t)$ is the density
operator ($c_{p,s}(t)$ is the electron annihilation operator).
For the depolarized spectrum, we just need to change $\chi_\rho$ to 
$\chi_\sigma$, the spin density correlation function, which is defined by
\begin{equation}
\chi_\sigma(q,t)=\langle\left[\sigma^\dagger(q,t),\sigma(q,0)\right]
\rangle_0,
\end{equation}
where $\sigma(q,t)\equiv\sum_{ps}s\,c^\dagger_{p+q,s}(t)c_{p,s}(t)$ is
the spin density operator.

Since $\chi_{\rho}(q,\omega)$ is equivalent~\cite{manhan} to the
reducible polarizability, $\Pi_{\rho}(q,\omega)$,
we may rewrite (Fig. 1) it as a sum of the series of irreducible polarizability,
$\Pi_{0,\rho}(q,\omega)$, bubble diagrams:
\begin{eqnarray}
\Pi_{\rho}(q,\omega)&=&
\Pi_{0,\rho}(q,\omega)+\Pi_{0,\rho}(q,\omega)V_c(q)
\Pi_{0,\rho}(q,\omega)+\cdot\cdot\cdot
\nonumber\\
&=&\frac{\Pi_{0,\rho}(q,\omega)}{1-V_c(q)\Pi_{0,\rho}(q,\omega)}
=\frac{\Pi_{0,\rho}(q,\omega)}{\epsilon(q,\omega)},
\end{eqnarray}
where $\epsilon(q,\omega)$ is the dynamic dielectric function.
For the spin density correlation function,
$\chi_\sigma(q,t)$, however, the equivalent
reducible response function, $\Pi_{\sigma}(q,\omega)$ is exactly the
same as the irreducible one, $\Pi_{0,\sigma}(q,\omega)$, due to
the spin independence of Coulomb interaction (i.e. $V_c$ is explicitly
spin independent). We
will then use RPA~\cite{manhan} 
to calculate $\Pi_{\rho}(q,\omega)$
and $\Pi_{0,\sigma}(q,\omega)$ and evaluate
the charge and spin density correlation function (reducible polarizability)
in the following sections.
\subsection{RPA and Hubbard approximation}
In the random phase approximation, which is very successful in the
high density interacting two- or three-dimensional electron systems (the 
corresponding Feynman diagrams are shown in Fig. 1), the
irreducible polarizability is approximated by the noninteracting 
electron-hole bubble (and therefore $\Pi_{0,\rho/\sigma}(q,\omega)=
\Pi^{\mathrm{RPA}}_{0,\rho/\sigma}(q,\omega)$).
We can calculate its 1D analytical expression at $T=0$ (we choose $\hbar=1$
throughout) to be
\begin{eqnarray}
\Pi^{\mathrm{RPA}}_{0,\rho/\sigma}(q,\omega)
&=&\frac{-2i}{(2\pi)^2}\int d\nu dp\; G_0(p,\nu)G_0(p+q,\nu+\omega)
\nonumber\\
&=&\frac{-1}{\pi}\int dp\
\frac{n_0(p)-n_0(p-q)}{\omega+i\delta-p^2/2m+(p-q)^2/2m}
\nonumber\\
&=&\frac{m}{\pi q}\ln\left[\frac{(\omega+i\delta)^2-(q^2/2m-qv_F)^2}
{(\omega+i\delta)^2-(q^2/2m+qv_F)^2}\right]
\nonumber\\
&=&\frac{m}{\pi q}\ln\left|\frac{\omega^2-(q^2/2m-qv_F)^2}
{\omega^2-(q^2/2m+qv_F)^2}\right|
\nonumber\\
&&+i\ \rm{sign}\it{}(\omega q)\left[
\theta\left( |\omega|-qv_F-\frac{q^{\rm 2}}{\rm{2}\it{m}} \right)
\right.\nonumber\\
&&-\left.
\theta\left( |\omega|-qv_F+\frac{q^{\rm 2}}{\rm{2}\it{m}}\right)\right],
\end{eqnarray}
where $G_0(p,\nu)$ is the bare electron Green's function and $n_0(p)=\theta
(k_F-|p|)$
is the noninteracting momentum distribution function.
The first term of the last equality is the real part while the second term
is the imaginary part of $\Pi^{\mathrm{RPA}}_{0,\rho/\sigma}(q,\omega)$,
from which we could determine the single particle
excitation region as shown in Fig. 2(a).
We should note that unlike the 2D or the 3D situation, the SPE continuum region
for 1D system is very restricted in the long wavelength limit ($q\ll k_F$).
In higher dimensions, the SPE continuum is
gapless in the low energy region
at any finite wavevector smaller than $2k_F$, but it is gapped in the
1D system
due to the phase space restriction. The strength of the SPE 
is a constant, independent of the frequency, $\omega$, as we
can see from Eq. (6),
so that its contribution must be very limited in the interacting
situation, which is dominated by the CDE or the 
collective plasmon excitations. From Eq. (6)
we also see how the SPE region disappears completely as
the band curvature goes to zero, i.e. when the band dispersion is
taken to be linear as in the LL model.

To improve the RPA calculation, which does not consider any vertex correction
in $\Pi^{\mathrm{RPA}}_{0,\rho/\sigma}(q,\omega)$, 
and therefore is exact only in the long
wavelength limit, Hubbard~\cite{manhan}
introduced a simple local field correction to RPA
by approximately summing the ladder exchange diagrams (and thus including
vertex correction in a very crude manner).  Instead of going through
the details of his derivation, we here only show the final 1D result, which is
simple~\cite{dyn_response} and can be applied to our calculation.
The RPA dielectric function
of Eq. (5) is replaced in the Hubbard approximation (HA) by
\begin{equation}
\epsilon_H(q,\omega)=1-
\frac{V_c(q)\Pi^{\mathrm{RPA}}_{0,\rho}(q,\omega)}
{1+V_c(q)G(q)\Pi^{\mathrm{RPA}}_{0,\rho}(q,\omega)},
\end{equation}
where the local field(vertex) correction, $G(q)$, 
in the 1D Hubbard approximation is given by~\cite{dyn_response}
\begin{equation}
G(q)=\frac{1}{2}\frac{V_c(\sqrt{q^2+k_F^2})}{V_c(q)}.
\end{equation}
In the long wavelength limit, $q\rightarrow 0$, $G(q)\rightarrow 0$ for
Coulomb interaction and the RPA
result is restored in the long wavelength limit as it must.
In Fig. 2 we plot the dispersion and spectrum
of the 1D charge density collective excitation (usually called the 1D plasmon
mode) within both the RPA and the Hubbard approximation
--- the plasmon mode is defined by the zero of the
dielectric function and the intensity or the spectral weight is
given by the imaginary part of the dielectric function
(i.e. the dynamical structure factor~\cite{manhan}).
\subsection{Charge density correlation function --- polarized spectrum}
Using the RPA result, Eqs. (5) and (6), we can obtain the
density correlation function as well as the polarized NRS spectrum:
\begin{eqnarray}
&&\mathrm{Im}\Pi_\rho(\it{q},\omega)=
\nonumber\\
&&\frac{\mathrm{Im\Pi}^{\mathrm{RPA}}_{0,\rho}(\it{q},\omega)}
{\left[\rm{1}-\it{V}_c(q)\rm{Re\Pi}^{\mathrm{RPA}}_{0,\rho}
(\it{q},\omega)\right]
^{\mathrm{2}}+
\left[\it{V}_c(q)\rm{Im\Pi}^{\mathrm{RPA}}_{0,\rho}(q,\omega)\right]^{\rm{2}}}.
\end{eqnarray}
Note that the imaginary part of $\Pi^{\mathrm{RPA}}_{0,\rho}$ 
is zero outside the SPE region.
Therefore we expect a delta function like excitation at
$\rm{1}-\it{V}_c(q)\rm{Re\Pi}^{\mathrm{RPA}}_{0,\rho}(\it{q},\omega)=\rm{0}$ for
the plasmon mode or the
CDE dispersion. The dispersion is plotted
in Fig. 2(a). At $T=0$ without any impurity scattering
the dispersion of the plasmon mode in RPA is given by
\begin{equation}
\omega_{pl}(q)=\sqrt{\frac{A(q)(q^2/2m+qv_F)^2-(q^2/2m-qv_F)^2}{A(q)-1}},
\end{equation}
where $A(q)=\exp[q\pi/mV_c(q)]$. In the long wavelength limit
($q\rightarrow 0$), the dispersion becomes
\begin{equation}
\omega_{pl}(q)=qv_{pl}=qv_F\sqrt{1+\frac{2}{\pi v_F}V_c(p)}+O(q^3),
\end{equation}
due to the fact that $V_c(q)\rightarrow\ln(1/|q|)$ diverges in the long
wavelength limit. Note that Eq. (11) is exactly the same as the dispersion
obtained in the pure LL theory, which is discussed in Sec. III and is 
shown in Eq. (15) below. In Fig. 2(a), we find that
the plasmon energy is actually larger than
the SPE continuum
energy for all momentum, so that there is no
Landau damping in the 1D system within the RPA calculation.
The 1D plasma dispersion has no gap in the long wavelength limit,
but an infinite slope at $q=0$ due to the logarithmic divergence
of the 1D Coulomb interaction.

In Fig. 2(b), we also show the typical calculated polarized RRS
spectrum (proportional to Im$\Pi$ in Eq. (9)) using a phenomenological 
broadening
factor, $\gamma=0.05 E_F$, which may be arising from impurity scattering. 
We find that
the spectral weight of CDE is much larger than that of SPE (about one
thousand times !). In the same figure we show the HA results as well.
We find that while the vertex
correction indeed increases the SPE weight somewhat relative to the
CDE weight,
the HA is still completely unable (by a factor of 100) to explain the
experimental finding~\cite{QWR_ref} 
of the SPE mode being comparable in the intensity
to the CDE mode (double-peak structure) in the polarized RRS 
spectrum~\cite{QWR_ref}.

Moreover, 
if the electron energy dispersion is linear (as it is close to 
the Fermi point), 
the SPE excitation spectral weight calculated by Eq. (9) disappears.
This indicates that the band
curvature around $k_F$ plays an important role
in forming the SPE peak in the experiments.
For example, in the linearized LL theory the SPE will have (Sec. III) exactly
zero spectral weight. In Sec. II-D below we consider the band nonparabolicity effect explicitly.
\subsection{Spin density correlation function --- depolarized spectrum}
As discussed above, 
the depolarized spectrum of 1D electron systems in the RPA
is just the imaginary part of
$\Pi^{\rm{RPA}}_{0,\sigma}(q,\omega)$ in Eq. (6), which is nothing but a square
function with maximum and minimum energies (for $\omega>0$), $qv_F+q^2/2m$
and $qv_F-q^2/2m$ respectively, the same as the SPE in the polarized spectrum.
Thus, within RPA the depolarized mode dispersion is identical to the SPE
energy, i.e. $\omega=qv_F$ with a
$q^2$ broadening.
One should note, however, that when vertex corrections such as in the
Hubbard approximation are taken into account,
the spectra of the spin density excitation (SDE) will not be
exactly the same as the SPE due to the vertex correction induced energy
shift. In Fig. 3 we show the SDE spectrum obtained by
calculating the spin density correlation function in RPA
and also in the Hubbard approximation. The
vertex correction shifts the SDE peak to lower energy (an excitonic shift)
in the HA and thus separates it from the SPE mode.
\subsection{The breakdown of momentum conservation and nonparabolicity effects
in RPA}
Beyond the standard RPA calculation, we include two
nongeneric effects, the breakdown of momentum-conservation (arising from 
impurity scattering, for example) and the
nonparabolicity of electron energy dispersion, because both of these 
corrections are likely to transfer some large wavevector SPE weight to
smaller wavevector. 
For the breakdown of momentum-conservation, 
we use a phenomenological approach~\cite{p_break} by
introducing a broadening function that couples the polarized spectrum 
$\chi_\rho(q,\omega)$ at momentum $q$ to that at momentum $q'$:
\begin{equation}
\chi_\rho(q,\omega;\Gamma)\sim\frac{\Gamma k_F}{\pi}\int dq'\,
\frac{\chi_\rho(q',\omega)}{(q-q')^2+\Gamma^2 k_F^2},
\end{equation}
where $\Gamma$ is a (phenomenological) 
dimensionless factor denoting the strength of the 
breakdown of 
momentum-conservation. For $\Gamma\rightarrow 0$ we get back the 
original spectrum. In Fig. 4(a), we show the numerical calculation results
of this effect by applying Eq. (12) onto the RPA result Eq. (9). At first
sight, one finds that finite $\Gamma$ does decrease the peak value of CDE
and enhance the SPE weight. For $\Gamma>0.5$, however, we find that the SPE 
peak merges into the very broad CDE peak, which is 
broadened also by the breakdown of momentum-conservation. In other words, the
breakdown of momentum-conservation reduces the CDE peak strength and also 
broadens
its width without changing either the total CDE spectral weight
or the SPE weight qualitatively. 
Therefore, in our direct numerical calculation,
we show that the breakdown of momentum-conservation (at least through
phenomenological treatments, such as Eq. (12)) is not the candidate
mechanism to provide an SPE spectral weight
comparable to the CDE weight in RPA calculations.

We now discuss the same issue 
by considering the band nonparabolicity effect
of the electron energy dispersion. We recalculate the RPA spectral weight 
including band nonparabolicity via an additional $q^4$ term
in the electron energy dispersion~\cite{wolff68}
\begin{equation}
\frac{E(q;\lambda)}{E_F}=\left(\frac{q}{k_F}\right)^2+
\lambda\left[\left(\frac{q}{k_F}\right)^4-\left(\frac{q}{k_F}\right)^2\right].
\end{equation} 
This expression of $E(q;\lambda)$ keeps the Fermi energy constant 
($E(k_F;\lambda)=E_F$) for
all $\lambda$ and changes the electron effective mass 
$m_e(\lambda)=m/(1-\lambda)$ consistently. In Fig. 4(b), we show the
calculated polarized RRS spectrum
for different values of $\lambda\leq 0.1$. We find that the enhancement of 
the SPE weight is very small, while the CDE peak almost keeps the same weight.
Using larger $\lambda$ will cause greater blue-shifts in both SPE and CDE 
energies due to the increase of Fermi velocity via Eq. (13), 
which disagrees with 
the experimental results (note that the 
standard RPA results provide very good agreement with the experimental results
in the excitation energies~\cite{QWR_ref}).
Therefore, the nonparabolicity
effect cannot enhance the SPE spectral weight to be comparable to
the CDE weight. 
\section{Luttinger liquid model}
The Luttinger liquid model~\cite{Haldane81,Review,schulzemery}
is thought to provide a generic low energy description for 1D electron
systems, which are
characterized by the LL
fixed point in the renormalization group sense. The standard and exactly
solvable LL model is the 1D electron gas with a linear dispersion 
($E(k)=rv_F(k-rk_F)$) around Fermi points ($\pm k_F$) at each branch 
($r=\pm 1$) and with short-ranged forward interaction~\cite{Haldane81,Review}. 
The Hamiltonian of this model is written as 
\begin{equation}
H=\sum_{rps} v_F(rp-k_F) :c^\dagger_{rps}c_{rps}:+H_4+H_2,
\end{equation}
where $:\cdot\cdot\cdot :$ is the normal order operator,
$H_{4}$ is for the electrons interacting within the right ($r=+1$) 
or the left ($r=-1$) band
only and $H_{2}$ is for the electrons interacting between these two bands.
Using bosonization method and a linear transformation, the Hamiltonian of
Eq. (14) can be exactly diagonalized by the two boson operators: 
charge boson $\rho_r(p)$ and spin boson $\sigma_r(p)$.
This fact makes the
collective excitation spectra (CDE and SDE) in the Luttinger model 
very simple: both the charge (CDE) and the spin (SDE)
modes are delta function
like poles and there is no SPE mode or equivalently, quasi-particle 
spectral weight, at all. In the
following, we think it is instructive, for the sake of completeness and
notational clarity, to present this result 
not only from the bosonization method we discussed above but also from 
the many-body Feynman
diagrammatic technique via Ward identity,
which is essentially equivalent to the RPA
calculation but with linear band dispersion.
\subsection{Bosonization method}
Evaluating the expectation value, we get for the
charge $(\rho)$ and the spin $(\sigma)$ sectors:
\begin{eqnarray}
\chi_\rho(q,\omega)&=&\frac{2L}{\pi}\frac{K_\rho(q)v_\rho(q)q^2}
{(\omega+i\delta)^2-(\omega^\rho_q)^2} \\
\chi_\sigma(q,\omega)&=&\frac{2L}{\pi}\frac{K_\sigma(q)v_\sigma(q)q^2}
{(\omega+i\delta)^2-(\omega^\sigma_q)^2},
\end{eqnarray}
where the excitation energy, $\omega^{\rho/\sigma}_q=|q|v_{\rho/\sigma}(q)=
|q|v_F/K_{\rho/\sigma}(q)$, and the charge sector 
Luttinger exponent $K_{\rho}(q)$ is 
\begin{equation}
K_\rho(q)=\left(1+\frac{2V_c(q)}{\pi v_F}\right)^{-1/2},
\end{equation}
while $K_{\sigma}(q)=1$ in the spin sector
for the spin-independent Coulomb interaction.

It is clear that the above results are completely spin-charge
separated, which is another important feature of the LL model.
One should note that there is no spectral weight in $\chi_\rho(q,\omega)$
at $\omega=qv_F$ for any SPE mode (or, for that matter any mode).
This shows that the small SPE peak
(compared to CDE) in the Fermi liquid-RPA theory is totally absent in 
the LL theory. Thus any possible explanation within the LL theory for the 
anomalous low energy peak in the polarized RRS spectra must arise 
from some mode (e.g. a multiboson mode or an SSE mode) 
other than the SPE mode which is completely absent 
in the LL theory.

We note one other aspect (the spin-charge separation mentioned above) 
of the LL theory in this context which has created some minor 
confusion. The spin-charge separation of the LL theory has nothing 
whatsoever to do with the separate existence of SDE/CDE in the 
depolarized/polarized RRS spectra. The collective spin and charge 
density excitations are completely distinct excitations in the FL 
theory as well --- they are the poles of the appropriate spin 
(for SDE) and charge (CDE) density correlation functions of the system which 
have totally different energies and selection rules (i.e. whether 
there is a spin-flip or not) in any reasonable theory. The reason 
spin-charge separation is rather complete in the LL theory is because 
the Luttinger liquid does not have any quasi-particles or single 
particle excitations --- it has only collective spin and charge 
excitations which are poles of different correlation functions and 
are always separate. Indeed, higher dimensional systems, such as 2D 
and 3D GaAs structures, exhibit qualitatively similar RRS spectra 
as in the 1D system with the CDE peak (and a weak SPE-like low 
energy feature) showing up in the polarized spectra and the SDE peak 
showing up in the (spin-flip) depolarized RRS spectra. These higher 
dimensional systems are obviously Fermi liquids and have no LL-like
intrinsic spin-charge separation while at the same time having 
distinct CDE and SDE collective modes.

Unlike the formulae for the single-particle Green's function,
in which the non-Fermi-liquid-like Luttinger liquid feature
arises from the nonperturbative
power-law behavior together with the velocity renormalization, 
the charge and spin
density correlation functions (which are two-particle Green's functions) have
no such power-law behavior at all. The density correlation functions and the
the associated charge/spin density excitation collective mode spectra are
essentially identical in the LL and the FL-RPA model
\cite{dyn_response,11} except for the complete
absence of any SPE spectral weight in the charge sector in the LL model.
The Luttinger liquid effects appear only
in the mode velocity renormalization, $v_\rho(q)$,
and the overall mode amplitude factors $K_{\rho}(q)$ in the collective
mode spectra.
\subsection{Diagrammatic method}
The diagrammatic method for the Luttinger liquid theory tells
us more
about the transition from the Fermi liquid to the Luttinger liquid, because
it is physically more transparent than the bosonization technique, which is
more of a formal mathematical tool. Early seminal
work by Dzyaloshinkii and Larkin~\cite{Larkin} and recent important work of
Schultz~\cite{ll_3kf} have shown that this method
is equivalent to the bosonization theory, even though its theoretical
structure follows a Fermi liquid type conventional many-body theory.

Eq. (5) is the starting point of the diagrammatic calculation.
In order to evaluate the
irreducible polarizability, we use the Ward identity connecting the Green's
function with the vertex function in the
following formula
\begin{equation}
\Gamma_{rs}(p,\nu,q,\omega)=\frac{G_{rs}^{-1}(p,\nu)-
G_{rs}^{-1}(p-q,\nu-\omega)}{\omega+i\delta-rqv_F},
\end{equation}
where $\Gamma_{rs}(p,\nu,q,\omega)$ is the vertex function of
two particle lines and one
interaction line. The Ward identity follows directly from the particle 
and current
conservation in each branch and spin (valid
only for forward scattering) coupled with linear dispersion relation.
It can be derived by summing the infinite series of vertex diagrams as
shown in Fig. 5. Using this vertex expression, one can calculate
the exact irreducible polarizability (consider the charge part only)
\begin{eqnarray}
\Pi_{0,\rho}(q,\omega)
&=&-i\sum_{rs}\int\frac{dp}{2\pi}\int\frac{d\nu}{2\pi}
G_{rs}(p,\nu)G_{rs}(p-q,\nu-\omega)
\nonumber\\
&&\hspace{1cm}\times\Gamma_{rs}(p,\nu,q,\omega)
\nonumber\\
&=&\sum_{rs}
\int\frac{dp}{2\pi}\frac{n_{rs}(p-q)-n_{rs}(p)}{\omega+i\delta-rqv_F}
\nonumber\\
&=&\frac{-2q/\pi}{\omega+i\delta-rqv_F}.
\end{eqnarray}
Comparing Eq. (19) with Eq. (6) we find that they are identical if
we only change the parabolic dispersion in Eq. (6) to the linear one as
in the Luttinger model. Therefore we obtain the striking result that
the irreducible
polarizability of the linear band dispersion model is exactly the same as
the RPA result. In other words, vertex corrections to the irreducible
polarizability vanish. This result can also
be verified by the topological argument given in ref.~\cite{ll_3kf}, which
shows that all the electron-hole loops connecting 
with more than three interaction lines cancel out.
Note that the above result
is independent of temperature, due to the particle number conservation
in the $p$-integral.

Using this formula of $\Pi_{0,\rho}(q,\omega)$, 
the charge density correlation function
can be easily obtained via Eq. (5). It is instructive, however, to see how
it works in Eq. (9). In the usual RPA calculation in 1D, the plasmon
mode (given by $1-V_c(q)\rm{Re}\Pi_{0,\rho}^{\rm{RPA}}=0$) 
is above the SPE region so that the square
function ($T=0$) of $\rm{Im}\Pi_{0,\rho}^{\rm{RPA}}(\it{q},\omega)$
still exists, unaffected by the CDE mode,
while the CDE itself is a delta function like pole because 
$\rm{Im}\Pi_{0,\rho}^{\rm{RPA}}(\it{q},\omega)=\rm{0}$ at the plasmon energy. 
When the dispersion is linearized as in the LL model,
$\rm{Im}\Pi_{0,\rho}^{\rm{RPA}}(\it{q},\omega)$ itself becomes a 
delta function at $\omega=qv_F$ rather
than a square function (see Eq. (19)), 
leading to the complete suppression and the
disappearance of the
SPE mode as we can see from Eq. (9) at $\omega=qv_F$.
This makes this diagrammatic result consistent with the bosonization result,
in which there is manifestly no single particle
eigenstate at all in the final spectrum. Both approaches predict the complete
absence of an SPE mode in the LL theory.

Finally we can conclude that the LL theory gives precisely the same collective
mode spectra as the corresponding RPA theory does with the only difference 
being the complete non-existence of any (even weak) SPE features in the
LL spectra.
\section{Hubbard model}
Motivated by the long-standing SPE-feature puzzle discussed 
in the Introduction, it has been 
suggested \cite{schulz93,sassetti} that we can interpret the 
"SPE" peak observed in the
experimental RRS spectra as the "singlet spin excitation" 
\cite{schulz93}.
In other words, the incident
photon virtually flips the electron spin and then restores its polarization
after the scattering, leaving the electron spin unchanged. Unlike the triplet
spin excitation, SDE, which manifests itself in the depolarized RRS spectra,
the virtual spin-flip process of SSE may, in principle, contribute to
the final spinless scattering matrix element of the polarized Raman 
scattering spectrum, so one could
expect that SSE should be very close to the SPE in energy under
the spin-independent Coulomb interaction. That explains why
one cannot simply separate these two (SSE and SPE) 
whether in the experimental measurement
or in the theoretical calculation. In the LL theory, there is no SPE, but
SSE is, in principle, allowed and may simulate the SPE of the FL-RPA theory.
To investigate the role of SSE in more details
we could use the 1D Hubbard model on a lattice, which 
can be mapped to
the exactly solvable Luttinger liquid model in the long wavelength limit.
Thus the 1D Hubbard model has no generic SPE properties, and could therefore
be useful in the understanding of SSE properties. We therefore study
the SSE in the 1D Hubbard model,
and investigate whether its spectral weight can be comparable to the
CDE as observed in the experiments.

In this paper, we want to study the 1D single band Hubbard model (HM) 
through the
Bethe-ansatz equations and the Lanczos-Gagliano method, which is shown to be in
excellent agreement with the exact diagonalization result.
Although HM is a lattice model of short-ranged on-site
interaction, unlike the realistic 1D QWR systems which are 
continuum systems with long-ranged Coulomb interaction, we could
still use it as a valid model in qualitatively discussing the problem 
of the polarized
Raman scattering because all 1D interacting systems belong to the LL universality class and generic issues may be addressed in any particular 1D model.
We first identify the
holon excitation of the Hubbard model to be the CDE,
and the triplet spinon to be the SDE in the usual RRS language, by
comparing their dispersion relations in the whole spectrum.
We then further obtain the finite spectral weights of SSE in the charge
density spectrum and show
that the weight of singlet spin density excitation is still rather low 
in the HM and
\textit{cannot} produce large
spectral weights in the polarized RRS scattering spectrum as 
found in the experiment.
Thus, our HM results shows that the "SPE" peak in the polarized
RRS experiments
is unlikely to be explained by the 1D singlet spin excitation,
at least within any nonresonant theory which considers the 
elementary excitations only in the conduction band.
\subsection{Theory}
The simple 1D single band Hubbard model,
\begin{equation}
H=-t\,\sum_{i,\sigma}\left(c^{\dagger}_{i+1,\sigma}c_{i,\sigma}+
\mathrm{H.c.}\right)+U\sum_{i}n_{i\uparrow}n_{i\downarrow},
\end{equation}
where $c_{i,\sigma}$ and $n_{i,\sigma}$ are respectively
the fermion creation operator and the density
operator for site $i$ and spin $\sigma$, has been extensively studied.
$t$ and $U$ are hopping energy and on-site short range
spin-dependent
interaction following
the usual notation in the literature~\cite{lieb,coll,yang}.
Note that the Hubbard model
is basically a model with a spin-dependent short range (on-site)
interaction, $U$.
It has generic LL properties in the long
wavelength limit and for low-lying excitation energy, i.e. no single particle
behavior in the spectral function at Fermi wavevector.
The explicitly spin-dependent
interaction, $U$, in the HM, however, should
make the spin singlet state more enhanced in the spectrum, and easier to study.
Among the many accurate and useful methods to study the 1D HM,
we use the Bethe-ansatz
method~\cite{lieb,yang}
to obtain the ground state and the low-lying excitation state dispersion
spectra. It is well-known that the Bethe-ansatz
wavefunctions are not particularly useful in calculating correlation functions,
and therefore we need an alternative method to obtain the spectral weights
of the elementary excitations.
We calculate the charge density and spin density correlation functions, to be
compared respectively with the polarized and depolarized 
inelastic light scattering spectra,
by using the Lanczos-Gagliano method
\cite{gagliono86,gagliono87,gagliono88,gagliono89}. 
This method (described and discussed below) gives a simple but
fast-convergent result for the correlation functions in the lattice model. 
By comparing the
momentum-energy dispersion relations of these two different methods,
(i.e. the Bethe-ansatz and the Lanczos-Gagliano method)
we can identify each important spectral peak obtained by the
Lanczos-Gagliano method to be a certain Bethe-ansatz elementary
excitation in the Hubbard model language (holon, triplet spin or singlet spin
excitations, for example) and then estimate their relative  weights
for comparison. Our results obtained by this technique are consistent 
with the quantum Monte Carlo
calculations~\cite{num_sc_separation} where appropriate.
\subsubsection{Bethe-ansatz equations}
It is well-known that the 1D Hubbard model can be solved exactly 
\cite{lieb,coll} by the
Bethe-ansatz method. The eigenvalue equation of Eq. (20) is proved to be
identical to solving the coupled system of equations (under periodic boundary
condition)
\begin{eqnarray}
&&\hspace{-3.5cm}
e^{ik_jL}=\prod^M_{\alpha=1}\frac{\sin k_j-\lambda_\alpha+iU/4}
{\sin k_j-\lambda_\alpha-iU/4}\\
\prod^N_{j=1}\frac{\lambda_\alpha-\sin k_j+iU/4}
{\lambda_\alpha-\sin k_j-iU/4}&=&-\prod^M_{\beta=1}
\frac{\lambda_\alpha-\lambda_\beta+iU/2}{\lambda_\alpha-\lambda_\beta-iU/2},
\end{eqnarray}
where $L(N)$ is the total number of sites(electrons) and $M$
is the number of down-spin electrons ($M\leq N/2$).
The pseudo-momentum, $\{k_j\}$,
and spin rapidities, $\{\lambda_\alpha\}$, are generally complex variables
to be solved and related to the physical states of energy, $E$, 
and momentum, $p$, by
\begin{equation}
E=-2t\,\sum^N_{j=1}\cos k_j,
\end{equation}
and
\begin{equation}
p=\sum^N_{j=1} k_j.
\end{equation}
If the $k_j$'s and $\lambda_\alpha$'s are all real, the identity
of the phases in Eqs. (21)-(23) can be obtained by taking the logarithm.
Then we have the following well known results,
\begin{equation}
Lk_j= 2\pi I_j+2\,\sum^{M}_{\alpha=1}\tan^{-1}
\left(\frac{\lambda_\alpha-\sin k_j}{U/4}\right)
\end{equation}
\begin{eqnarray}
2\,\sum^{N}_{j=1}\tan^{-1}\left(\frac{\lambda_\alpha-\sin k_j}{U/4}\right)
&=&2\pi J_\alpha 
\nonumber\\
&&\hspace{-1cm}+2\,\sum^{M}_{\beta=1}\tan^{-1}\left(
\frac{\lambda_\alpha-\lambda_\beta}{U/2}\right)
\end{eqnarray}
where the quantum numbers, $\{I_j\}$, are all distinct from each other and are
integers if $M$ is even and are half-odd integers if $M$ is odd, and are only
defined in $|I_j|\leq L$. Similarly, the set $\{J_\alpha\}$ are
all distinct and
are integers if $N-M$ is odd and half-odd integers if $N-M$ is even. The value
of $\{J_\alpha\}$ is restricted by $|J_\alpha|<(N-M+1)/2$. Generally, it is
not hard to use the Bethe-ansatz equations to solve large size systems.
In the thermodynamic limit, $L\rightarrow\infty$, one can find the
equivalent linear integral equations for the density of $k$'s and
$\lambda$'s on the real axis~\cite{coll,woynarovich}. But we will only
focus here on the finite size
systems in order to compare the Bethe-ansatz results with
the results of the Lanczos-Gagliano
method, which is necessarily computationally restricted to small system sizes.

To solve these Bethe-ansatz equations, we first have to define the
proper quantum numbers, $\{I_j\}$ and $\{J_\alpha\}$; 
then solve Eqs. (25)-(26) to get $k_j$'s
and $\lambda_\alpha$'s, and then get the
momentum, $p$, and the energy, $E$, of that state specified by those quantum
numbers. Here we present the quantum number structures of the ground state and
two low lying excited state, the "$4k_F$" singlet states, 
the "$2k_F$" triplet states, and the "$2k_F$" singlet states as first 
named by Schultz~\cite{schulzemery}. The first two
have $k$'s and $\lambda$'s all real in Eqs. (25)-(26), while the last one
has one pair of complex $\lambda$'s in Eqs. (23)-(24)

\underline{\textbf{Ground state:}}
It is easy to see that the ground state is nondegenerate
only if $N$ is of the form $4m+2$ (m is an integer). In the following,
we just study the nondegenerate case for simplicity. Considering the
essential symmetries, one can write the ground state
quantum number satisfying the above restrictions to be
\begin{eqnarray}
\{I_j\} &=& \{-(N-1)/2,\cdot\cdot\cdot,(N-1)/2\},\\
\{J_\alpha\} &=& \{-(N/2-1)/2,\cdot\cdot\cdot,(N/2-1)/2\}.
\end{eqnarray}

\underline{\textbf{"$4k_F$" singlet state (holon excitation):}}
The first simplest excited states are obtained by removing one of the momentum
quantum numbers, $-(N-1)/2+i_0$, in $\{I_j\}$ and adding a "new" one at 
$(N-1)/2+I_0$
outside the ground state sequence. All other momentum quantum numbers and spin
quantum numbers are kept the same as in the ground state structure. Therefore,
the new sequence of $\{I_j\}$ is
\begin{eqnarray}
\{I_j\}&=&\{-(N-1)/2,\cdot\cdot\cdot,-(N-1)/2+i_0-1, \nonumber \\
& &-(N-1)/2+i_0+1,\cdot\cdot\cdot,(N-1)/2, \nonumber \\
& &(N-1)/2+I_0\},
\end{eqnarray}
and the $\{J_\alpha\}$ is the same as the ground state in Eq. (28).
Therefore, there are
two free parameters, $i_0$ and $I_0$, for this type of excitations.
In this paper, we use $(i_0, I_0)$ to denote this excitation state.
According to Schultz~\cite{schulzemery}, 
they are named "$4k_F$" singlet states due to their
energy minimum at $k=4k_F$ in their dispersion spectrum. In the
literature, these states are also called "particle-hole excitation"
or "holon" excitation~\cite{coll}.
In the rest of this paper, we will call them "holon" excitations
for simplicity. (This is related to the CDE of our earlier sections.)

\underline{\textbf{"$2k_F$" triplet state (triplet spinon excitation):}}
Next we consider the excitations of the $J$'s with all $\lambda$'s and $k$'s
real. This is possible only if $M<N/2$.
The simplest excitations of this type are
obtained by considering $M=N/2-1$.
The total spin of the system is $S=1$, so we expect this
excitation to be related to a triplet
spin excitation. The quantum numbers of
these states are 
\begin{eqnarray}
\{I_j\} &=& \{-N/2+1,\cdot\cdot\cdot,-N/2+i_0-1, \nonumber \\
&&-N/2+i_0+1,\cdot\cdot\cdot,N/2,N/2+I_0\},
\nonumber \\
J_1 &=& -N/4+\delta_{\alpha_1,1}, \nonumber \\
J_\alpha &=& J_{\alpha-1}+1+\delta_{\alpha,\alpha_1}+\delta_{\alpha,\alpha_2},
\end{eqnarray}
where $\alpha=2,...,M$ and $1\leq\alpha_1<\alpha_2\le M+2$. Here $\alpha_1$ and
$\alpha_2$ are the free parameters in the spin quantum number, $\{J_\alpha\}$,
and
$i_0$ and $I_0$ are the two parameters in momentum quantum number, $\{I_j\}$.
From Eqs. (23)-(26), we can see that $i_0$ and $I_0$ shift the total
momentum and energy of the spectrum created by the spin excitation in
$\{J_\alpha\}$. In the following calculation, we use
($i_0,I_0,\alpha_1,\alpha_2$) to denote these states in the spectrum.
These excitations are called "$2k_F$"
triplet states because its minimum energy is at $k=2k_F$.
In the rest of
this paper, we will call them "triplet spinon" for simplicity.
(This is related to the SDE of the earlier sections.)

\underline{\textbf{"$2k_F$" singlet state (singlet spinon excitation).}}
The third possible elementary excitations are from the complex solutions of
Bethe-ansatz equations, Eqs. (21)-(22).
Spin singlet states ($M=N/2$ and then
$S=0$) are obtained by having one pair of the complex conjugate,
$\lambda_\pm=\lambda_R\pm\lambda_I$ with all
other $k$'s and $\lambda$'s real. The new set of Bethe-ansatz equations are
obtained to be
\begin{eqnarray}
Lk_j &=& 2\pi I_j+2\sum^{M}_{\alpha\neq\alpha_1,\alpha_2}\tan^{-1}
\left(\frac{\lambda_\alpha-\sin k_j}{U/4}\right)
\nonumber\\
&+&2\left[\tan^{-1}\left(\frac{\lambda_R-\sin k_j}
{U/4-\lambda_I}\right)+
\tan^{-1}\left(\frac{\lambda_R-\sin k_j}{U/4+\lambda_I}\right)\right]
\nonumber\\
\\
&&\hspace{-1cm} 
2\sum^{N}_{j=1}\tan^{-1}\left(\frac{\lambda_\alpha-\sin k_j}{U/4}\right)=
2\pi J_\alpha
\nonumber\\
&&\hspace{-1cm}+2\sum^{M}_{\beta\neq\alpha_1,\alpha_2}\tan^{-1}
\left(\frac{\lambda_\alpha-\lambda_\beta}{U/2}\right)
\nonumber\\
&&\hspace{-1cm}+2\left[\tan^{-1}\left(\frac{\lambda_\alpha-\lambda_R}
{U/2-\lambda_I}\right)+
\tan^{-1}\left(\frac{\lambda_\alpha-\lambda_R}{U/2+\lambda_I}\right)\right],
\end{eqnarray}
where $j=1,2,...,N$ and $\alpha = 1,2,...,M$, but
$\alpha\neq\alpha_1,\ \alpha_2$.
The two equations for the complex $\lambda_\pm$ are
\begin{eqnarray}
&&\frac{1}{2}\sum^N_{j=1}\log\left(\frac{(\lambda_R-\sin k_j)^2
+(U/4+\lambda_I)^2}{(\lambda_R-\sin k_j)^2+(U/4-\lambda_I)^2}\right)
\nonumber\\
&&=\frac{1}{2}\sum^M_{\beta\neq\alpha_1,\alpha_2}\log
\left(\frac{(\lambda_R-\lambda_\beta)^2+(U/2+\lambda_I)^2}
{(\lambda_R-\lambda_\beta)^2+(U/2-\lambda_I)^2}\right) 
\nonumber\\
&&+\log\left(\left|\frac{4\lambda_I+U}{4\lambda_I-U}
\right|\right)
\\
&&\sum^{N}_{j=1}
\left[\tan^{-1}\left(\frac{\lambda_R-\sin k_j}{U/4+\lambda_I}\right)+
\tan^{-1}\left(\frac{\lambda_R-\sin k_j}{U/4-\lambda_I}\right)\right]
\nonumber\\
&&=2\pi J
\nonumber\\
&&
+\sum^{M}_{\beta\neq\alpha_1,\alpha_2}\left[
\tan^{-1}\left(\frac{\lambda_R-\lambda_\beta}{U/2+\lambda_I}\right)+
\tan^{-1}\left(\frac{\lambda_R-\lambda_\beta}{U/2-\lambda_I}\right)\right].
\nonumber\\
\end{eqnarray}
where
\begin{equation}
J=\left\{ \begin{array}{ll}
  \mbox{integer} & \hspace{-1.5cm}
             \left\{\begin{array}{ll} \mbox{if $|\lambda_I|>U/4$,
                                                   and $N-M$ is even}\\
                            \mbox{or if $|\lambda_I|<U/4$, and $N-M$ is odd}\\
                          \end{array}
                   \right.\\
  \mbox{half odd integer} & \mbox{otherwise.}
  \end{array}
\right.
\end{equation}
As for the quantum number, $\{I_j\}$ and $\{J_\alpha\}$, in the singlet states,
we choose them to be the same as the ground state, Eqs. (27)-(28), except for
the two free "holes" at $J_{\alpha_1}$ and $J_{\alpha_2}$, whose related spin
quantum numbers, $\lambda_{\alpha_1}$ and $\lambda_{\alpha_2}$ are replaced by
the pair of complex conjugate, $\lambda_\pm=\lambda_R\pm\lambda_I$. Eqs.
(33)-(34) are usually too complex to give a nontrivial solution because the
usual numerical iteration method will converge to the trivial
$\lambda_I=0$ solution. But one can simplify these equations by
taking $\lambda_I=U/4$ and one $k_j$ satisfying
$\sin k_j=\lambda_R$, so that Eq. (33) could be neglected, and all the terms
containing $\tan^{-1}[(\lambda_R-\sin k_j)/(U/4-\lambda_I)]$ in Eq. (34)
contribute a
phase $\pm\pi$. The phase number, $J$, is set to make the total phase shift
(including those from $\tan^{-1}[(\lambda_R-\sin k_j)/(U/4-\lambda_I)]$) to be
zero in the calculation.

The spin singlet excitations have a dispersion similar to the triplet ones.
Here we could use $(\alpha_1,\alpha_2)$ as the quantum number to define
these states. In the finite size system with repulsive interaction, $U$,
the singlet states have higher energy than
the triplet ones, but they will become degenerate in energy as we go to the
thermodynamic limit ($L\rightarrow\infty$, and $\langle n
\rangle$=constant). In the experiment, the spin triplet excitations 
(i.e. SDE) are
observed in the depolarized RRS spectra where a net spin-flip occurs,
while the singlet states are observed in the polarized spectra, which involve
no net spin-flip.

As mentioned in the beginning of this section,
the Bethe-ansatz method does not, in general, provide the spectral weights
for their solutions. Therefore, the three
elementary excitations above may not be equally important from the experimental
point of view, i.e. they may carry very different spectral weights (and some
may even be unobservable in the experimental spectra). 
All we know from the Bethe-ansatz solutions are the existence and the dispersion
of these excitations but $not$ their spectral weights. 
We also know that these are allowed solutions of the HM just as the SPE is 
an allowed solution of the FL-RPA model (but \textit{not} the LL model).
Comparing the mode spectral weights calculated by Lanczos-Gagliano
method we discuss next,
we calculate the relative spectral weights of these solutions and
then study their interaction dependence.
\subsubsection{Lanczos-Gagliano method}
In this paper Lanczos-Gagliano method means the combination of two
important
techniques in the lattice model. The standard Lanczos method
is to construct an $L\times L$ matrix representation for the tridiagonal
Hamiltonian,
like Eq. (20), and then directly diagonalize it to 
get the eigenvalues, $E_n$, and eigenfunctions,
$\Phi_n$, which could be used to do further
calculations, such as obtaining spectral weights.
But since only ground
state energy and wave function are needed in calculating the correlation
function by using the Gagliano's method (see below), we use a simpler but more
efficient way, the modified Lanczos method,
to calculate the ground state energy and
wave function. This method has been analyzed and discussed in detail
in references~\cite{gagliono86,gagliono87,gagliono88,gagliono89}, and
here we only provide a brief
review for the sake of completeness.

In the modified Lanczos method, we first present $H$ in a $2\times 2$
matrix in the basis of $\psi_0$ and $\psi_1$, which are both normalized and
orthogonal with each other by setting
\begin{equation}
\psi_1=\frac{H\psi_0-\langle H\rangle\psi_0}{(\langle H^2\rangle-
\langle H\rangle^2)^{1/2}}.
\end{equation}
where $\psi_0$ could be an
arbitrary trial wave function, and $\langle H^n\rangle
\equiv\langle\psi_0|H^n|\psi_0\rangle$. After diagonalizing the $2\times 2$
matrix representation of $H$, we can
get an eigenenergy, $\varepsilon_0$, and a wave function,
$\tilde\psi_0$ (subscript $0$ means the lower energy eigenstate).
This $\varepsilon_0$ and $\tilde\psi_0$ are then a better
approximation to $E_0$ and $\Phi_0$ than the initial trial state,
$\langle H\rangle$ and $\psi_0$. Using the same process by
iterationally replacing $\psi_0$ by $\tilde\psi_0$ to construct a new
$2\times 2$ matrix representation for $H$, we will then get a more accurate
approximation to the ground state energy and wave function simply
by diagonalizing the $2\times 2$ matrix.

The Gagliano's method~\cite{gagliono86,gagliono87,gagliono88,gagliono89}
is to use an infinite continued fraction representation
to calculate a general dynamic correlation function, $\chi_A(q,\omega)$, where
$A$ could be any operator. 
Its advantage is that one just needs
the ground state information to get the dynamical results extremely efficiently
in a very simple
and straightforward way. The convergence and accuracy of the calculation
is very good compared with the much more time consuming exact
diagonalization method~\cite{gagliono88}. 
In this paper, we
want to study the charge density correlation function, $\chi_\rho(q,\omega)$,
and the spin density correlation function, $\chi_\sigma(q,\omega)$.
Their corresponding operators, $A$'s, are respectively defined as
\begin{equation}
A(q) = \rho(q)\equiv\frac{1}{\sqrt{N}}\sum^{L}_{l,\sigma}e^{-iql}
\left(c^{\dagger}_{l,\sigma}c_{l,\sigma}-\langle n\rangle\right),
\end{equation}
and
\begin{equation}
A(q) = \sigma(q)\equiv
\frac{1}{\sqrt{N}}\sum^{L}_{l,\sigma}e^{-iql}
\sigma c^{\dagger}_{l,\sigma}c_{l,\sigma},
\end{equation}
where $\langle n\rangle=N/L$ is the average density.
The Gagliano's method introduces a function, $G_A(Z)$, in the following
formula to calculate the correlation function
\begin{eqnarray}
\chi_{A}(q,\omega)&=&\sum_{n}\langle\Phi_0|A^{\dagger}(q)|\Phi_n\rangle
\langle\Phi_n|A(q)|\Phi_0\rangle \nonumber \\
& &\times\delta(\omega-(E_n-E_0))\\
&=&\frac{1}{\pi}\,\mathrm{Im}\mathit{G_A}(\omega+E_{\mathrm{0}}+i\gamma),
\end{eqnarray}
where $\gamma$ is a (small) phenomenological 
broadening factor, and the function, $G_A$ is given by
\begin{eqnarray}
G_A(Z)&=&\langle\Phi_0|A^{\dagger}(q)(Z-H)^{-1}A(q)|\Phi_0\rangle \nonumber \\
&=&\frac{\langle\Phi_0|A^{\dagger}(q)A(q)|\Phi_0\rangle}
{Z-a_0-\frac{\textstyle b^2_1}{\textstyle Z-a_1-\frac{\textstyle b^2_2}
{\textstyle Z-\cdot\cdot\cdot} } }.
\end{eqnarray}
The coefficients, $\{a_n\}$ and $\{b_n\}$ are defined by
generating a set of orthogonal states with the following iteration formulae,
\begin{eqnarray}
|f_0\rangle &=& A(q)|\Phi_0\rangle \nonumber \\
|f_{n+1}\rangle &=& H|f_n\rangle-a_n|f_n\rangle-b^2_n|f_{n-1}\rangle\\
a_n &=& \langle f_n|H|f_n\rangle / \langle f_n|f_n \rangle \\
b^2_{n+1} &=& \langle f_{n+1}|f_{n+1} \rangle / \langle f_n|f_n \rangle,\ \
\mathrm{and}\ \ \it{b}_{\rm 0}=0.
\end{eqnarray}
Therefore, after we get $E_0$ and $\Phi_0$ from the modified Lanczos method,
we can calculate $\{a_n\}$ and $\{b_n\}$ iteratively and then obtain the
correlation function $\chi_\rho(q,\omega)$ and $\chi_\sigma(q,\omega)$ by 
using Eqs. (39)-(44).
Our calculation of the expectation value is done in the zero temperature 
ground state of the system.
The method may be easily generalized to finite temperatures~\cite{gagliono89},
which is, however, unnecessary for our purpose of studying the 1D excitation
spectra.
\subsection{Results and discussion}
We study the 1D Hubbard chain with three different densities,
$\langle n\rangle=N/L=1/3$ for 6 electrons in 18 sites, $\langle n\rangle=1/2$
for 6 electrons in 12 sites, and $\langle n\rangle=5/6$ for 10 electrons
in 12 sites. (Note that the usual filling factor of the system is 
$\langle n\rangle/2$ since our definition of density does not include spin.)
The size of the Hubbard chain is dictated here entirely by
computer memory restrictions in
calculating the correlation function via the Lanczos-Gagliano
method. We keep the electron
number to be $4m+2$ with $m$ an integer in order to have a nondegenerate
and zero-momentum ground state under the periodic boundary condition.
Throughout our calculations, we set the broadening factor, $\gamma$ in
Eq. (40), to be 0.01$t$ (where $t$ is the nearest-neighbor hopping
amplitude in Eq. (20)) and use the modified 
Lanczos method to calculate the ground
state energy iterationally until convergence to within less than $0.1\%$
in the ground state energy is reached. We
also truncate the infinite continuous fraction
in Eq. (41) at 25-27th order terms, which gives us good convergent
results in the calculation.

In the following, we will first discuss the results related to the polarized
spectrum, which involves no net spin-flip in the system, 
by using the two
methods mentioned above at a fixed interaction strength, $U/t=3$.
Then we consider the depolarized spectrum under the same conditions.
Finally we discuss their interaction, $U$, dependence by varying $U/t$
in our calculations. In the discussion below the terms "resonance"
or "resonance peaks" refer to the Lanczos-Gagliano calculations.
\subsubsection{Polarized spectrum analysis}
We compare the
dispersion of the charge density excitation with the
dispersions of the "$4k_F$" singlet states (holon) and the
"$2k_F$" singlet states
(singlet spinon) given by the solutions of the Bethe-ansatz equations, because
these two are the low lying elementary spinless ($S=0$) excitations
of the 1D Hubbard model and as such should correspond to the
polarized spectrum. We will also
study their relative spectral weights.
Both lower density ($\langle n\rangle =1/3$) and higher
density ($\langle n\rangle=5/6$) results are shown together (Fig. 6-9)
for further
discussion.

In Fig. 6(a), we show the spectral dispersion obtained by the
poles of the imaginary part of the charge density correlation
function. The doping density is $\langle n\rangle=1/3$
for 6 electrons in 18 sites in the 1D Hubbard chain with periodic boundary
conditions.
The center of each open diamond represents the position of the pole, and
its area is proportional to the spectral weight of that excitation.
In the same figure,
the dispersions of the holon (star) and singlet spinon (open square)
excitations given by the
solutions of Bethe-ansatz equations are also shown for comparison.
Several features could be noted from Fig. 6:
(i) the excitation peaks of the charge density
correlation function have a linear dispersion in the long wavelength
limit ($q\ll k_F=\pi/6$) and its slope gives the velocity of charge density
excitation of 1D Hubbard model.
(ii) The resonance peaks form a wing upto the
large momentum region (i.e. low energy excitations correspond
to the low momentum and high ones to high momentum), with a maximum
energy $\omega=4t$ at $q=\pi$. (iii)
The sizes of the diamonds, which represent their spectral weights, show
that the peaks at higher energy generally have
greater spectral weights than the ones at lower energies at the same momentum
(i.e. spectral weights are greater for peaks at higher energies).
Therefore one could see by eye a sine-like curve at the upper edge of the
resonance wing with a maximum at $\omega=4t$, and this observation
is consistent with the
results from quantum Monte Carlo simulations on larger
system~\cite{num_sc_separation}. (iv) There
are no excitation states for singlet spinons at small $q=\pi/9$. This implies
that the singlet spinon of the 1D Hubbard model is not allowed
for momentum smaller
than $2\times 2\pi/L$, where $2\pi/L$ is the momentum scale of this
finite size ($L$) system. This follows from the fact that the singlet
state must be excited by a pair of complex conjugate $\lambda_\pm$ in
Eqs. (21)-(22), which is at least a two-particle excitation, so that the
minimum momentum required is $2\times 2\pi/L$. (v)
One interesting feature is that there are clearly two energy minima at
$q=2k_F=\pi/3$ and $q=4k_F=2\pi/3$ in the spectrum. Comparing these
resonance peaks with the solutions given by Bethe-ansatz equations,
we find that the holon excitations cover almost exactly the same region
including the energy minimum at $4k_F$ except for the lower-lying peaks
around $2k_F$, where the singlet spinon just matches those peaks.
In other words, we could reasonably claim that the most dominant
features of the resonance peaks given by the charge density correlation
function arise from a combination of holon and singlet
spinon excitations in the 1D Hubbard model.
This result could not be trivially obtained
either by solving Bethe-ansatz equations or by calculating the charge density
correlation function alone,
as mentioned in the introduction --- one must combine
the two techniques to come to this conclusion.
Other spin singlet excitations given by the
solutions of Bethe-ansatz equations (for example,
2 pairs of complex $\lambda$'s
in Eqs. (21)-(22)), carry very small
spectral weights because no other
significant resonance peaks are found in this
dispersion spectra, except for some trivial ones. In the thermodynamic
limit, we expect that only the $4k_F$ holon and $2k_F$ singlet spinon
will have finite spectral weights and could be interpreted as the
"charge density excitation" and "single particle excitation" 
in the RRS spectra respectively
when comparing with the experiments~\cite{QWR_ref} as we mentioned in
the earlier sections. We
discuss this issue further later in this paper.

In Fig. 6(b), we show the imaginary part of the charge density correlation
function of the same system at $q=2\pi/9$. It shows
that singlet spinons have a relatively small, but non-negligible weight,
compared
with the weight of the dominant holon excitations.
Their relative spectral weight ratio (singlet spinon/holon) is less than 0.1.
Similar results are also obtained in the
systems of 6 electrons in 12 sites, $\langle n\rangle=1/2$, which are
not shown in this paper.

In Fig. 7(a), we show the dispersion of the resonant poles of the
charge density correlation function of 10 electrons in 12 sites
($\langle n\rangle=5/6$).
Here the holon excitations form a more narrow wing than in the
lower density system, but the basic shape of the dominant curve is
almost the same. Below this curve, the singlet spinon occupies almost
the whole resonance region. Since $4k_F=5\pi/3$ in this high density system,
we cannot see the $4k_F$ energy minimum in this figure (actually, the gap of
this energy minimum is very large in this finite system, but will go to zero
in the thermodynamic limit~\cite{schulzemery}).
But one could still see the energy
minimum of the singlet spinon at $2k_F=5\pi/6\sim 0.833\pi$ in Fig. 7(a).
There is a notable feature in these results at energies
above the holon excitations: there are some additional states, having
an energy gap equal to $U(=3t)$ and a maximum energy
greater than $4t$ at $q=\pi$ in this high density situation.
These states (solitons) must arise from
the double occupancy
of the electrons in the Hubbard model, which consequently explains their
high energy status. Quantum Monte Carlo method, to
the best of our knowledge, does not provide any information about these
high energy double occupancy states at the same density $n=5/6$
in the literature~\cite{num_sc_separation}.
From the Bethe-ansatz equations point of view, however,
these states should be obtained by taking the complex momentum,
{$k_j$}, solutions in Eqs. (21)-(22)~\cite{woynarovich},
whether in the high or the low density system.
Once again, we see the importance of studying the spectral
weights of the Bethe-ansatz solutions by comparing them to the correlation
function results so that one could tell the most realistic and physically
meaningful states. Just having the solutions, without much idea about their
spectral weights, is not useful in determining the experimental and/or
physical relevance of the particular excitations.
Therefore, as shown in Fig. 7(b),
the three most important contributions to the charge
density correlator arise from
singlet spinon, holon, and soliton excitations (double occupancy
excitations), from
lower energy to higher energy regime respectively in the 1D Hubbard chain.
Their relative
spectral weights show that the singlet spinon has the smallest weight,
and it could be shown that there are \textit{no} gapless holon and singlet
spinon excitations in the half-filling ($\langle n\rangle=1$) 
1D Hubbard model systems,
where the soliton (double occupancy excitations) and other higher
energy states
dominate the excitation spectrum.
\subsubsection{Depolarized spectrum analysis}
In Fig. 8(a), we show the resonance dispersion spectrum of the spin density
correlation function, $\langle\sigma\sigma\rangle$,
of the low density system (6 electrons
in 18 sites). The triplet spinon excitation spectrum given by the solutions of
Bethe-ansatz equations is also presented for comparison. Several
features are found: (i) in the long wavelength limit, the resonant poles have
a linear dispersion, whose slope gives the velocity of the
spin density excitation.
One could easily see that this velocity is always smaller than the velocity of
the charge density excitations at the same density. This is consistent with
result of previous work~\cite{schulzemery}.
(ii) The resonance poles form a
wing upto the large momentum region ($q\sim\pi$), whose maximum excitation
energy is below $4t$. 
(iii) Unlike the results for the charge density
correlation function in Fig. 8(a),
the most dominant poles are located in the lower energy part of the resonance
wing, which correspond to the triplet state without 
any excitations in $\{I_j\}$, and
therefore is related to the lowest energy ones in our calculation.
(iv) The resonance spectrum has an energy minimum at $2k_F$.
Compared to the triplet solutions of Bethe-ansatz equations,
the triplet spinon excitation spectrum has only three peaks matching
the resonance poles of the largest spectral weight at their momentum values
($q=\pi/9,
2\pi/9,$ and $3\pi/9$), and the other three match the poles of relatively much
\textit{weaker} states (not visible in Fig. 8(a), but distinguishable
in the absorption spectrum shown in Fig. 8(b)).
This result demonstrates that the spectral weights of elementary
excitations could be very different even if they result from the same
type of the Bethe-ansatz solution.
Fig. 9 shows the dispersion relation
of the spin density correlator of the large density system
($\langle n\rangle=5/6$) and the corresponding triplet spinon excitations
by the Bethe-ansatz solutions are also shown.
\subsubsection{Interaction dependence}
In this section, focusing on the lower density system ($L=18$ and $N=6$) and
a fixed momentum ($k=2\pi/9$), we study the mode
dispersion and spectral weight
of these excitations in a range of finite interaction ($U/t\le 10$)
to obtain the interaction dependence of the excitation spectra.
First, we study the polarized spectrum
given by the imaginary part of the charge
density correlation function, $\langle\rho\rho\rangle$ (shown in Fig. 10(a)).
Then we compare the
energy of the resonance peaks in the series of spectra with the Bethe-ansatz
results (Fig. 10(b)), and discuss the interaction dependence of the mode
velocity (Fig. 10(c)).
Finally we discuss the interaction
dependence of the spectral weight for each elementary excitation
(Fig. 11).

In Fig. 10(a), there are basically three peaks in the typical structure of the
polarized spectrum, and we can identify them as the singlet spinon, the second
and the first holon excitation (from lower to higher energy)
by explicitly comparing
with the energy given by the Bethe-ansatz solution in Fig. 6(a). Using
the notation introduced in Sec. IV-A1, the singlet
spinon is the state (2,3), while the holon I and II states are (6,2) and (5,1)
respectively. Several interesting features can be found in Fig. 10(a):
(i) in the noninteracting ($U=0$) case, there are only
two equal weight poles, which could be understood as the
two single particle (electron and hole) excitations around Fermi
surface, $k=k_F$. (ii) When finite interaction is turned on, there is
an additional excitation.
According to the comparison of dispersion relations, both
the new peak and the higher energy peak should be interpreted as holon
excitations (called holon II and holon I respectively, corresponding to
different $\{I_j\}$'s). (iii) The singlet spinon excitation
(shown in Fig. 10(a))
has a rapidly decreasing spectral weight with increasing
interaction, and disappears totally as $U/t>5.0$. (iv) The two holon
excitations shift to higher energy as $U$ increases, and maintain almost
the same spectral weight except one more peak appears as $U/t\geq8.0$
(see Fig. 10(a)). Above $U/t=8.0$,
the appearance of the new small peak affects
both the spectral weight and excitation energy of the holon II excitations
(see Figs. 10(b) and 11(a)). There are basically two possible interpretations
for this result. One is that this peak does not represent real
excitations, but may be arising spuriously from the inaccuracy of
the finite truncated
series or finite iteration used in the 
Lanczos-Gagliano method in the large $U$ range
(see Eq. (41) and the
discussion in the last section). Another possible reason is that it 
may arise from
the higher energy excitation states of unknown origin,
which are also
obtainable from the Bethe-ansatz solutions, but whose
strength is visible only when the
interaction strength is large enough.
We will not discuss this anomalous peak any further in this paper since
this falls outside the scope of our main interest.

In Fig. 10(c), we plot the excitation velocity, which is defined to be
\begin{equation}
v\equiv\left.\frac{\Delta E(q)}{\Delta q}\right\rfloor_{q\rightarrow 0^+},
\end{equation}
as a function of interaction strength. We find that when the interaction is
weak ($U/t\leq 1$), the two (holon and singlet spinon) excitations
are almost degenerate, while their
relative spectral weights change a lot (see Fig. 11(a)) as a function
of $U/t$. When $U/t$ increases,
the holon has greater excitation energy and hence velocity,
but the velocity of the singlet spinon decreases fast. This result holds
even in the thermodynamic limit.

In Fig. 10(b), we see more clearly that the energies
of the three elementary excitations are only weakly dependent
on the interaction
for small $U$, but strongly dependent on $U$ for large $U$.
In Fig. 11(b), we have a logarithmic scale 
in the spectral weight dependence with
respect to the interaction in small $U/t$ range ($U/t<1.0$).
By calculating the slope of these data, we find that the spectral
weight of holon I, $S_{\mathrm{holon\ I}}$, is almost a constant in the
smaller interaction range ($U<0.5t$), and then weakly decreases for higher $U$.
However, the spectral weight of holon II has a stronger power-law behavior,
$S_{\mathrm{holon\ II}}\propto U^{1.635}$.
Thus the two holon excitations differ a great deal in their
interaction dependence of their respective spectral weights.

In Fig. 12(a) we show the calculated depolarized spectra by taking
the imaginary part of spin density correlation function,
$\langle\sigma\sigma\rangle$ for various interaction strengths.
In the noninteracting case,
the spectrum is the same as the polarized one in Fig.
6(a) due to spin rotational symmetry. But with increasing interaction
strength both triplet spinon peaks 
move to lower energy in contrast to the polarized spectra.
Compared with the Bethe-ansatz
results in Fig. 12(b), the lower/higher energy peak, triplet spinon I/II,
is the state denoted by (6,0,1,3)/(6,1,2,3).
The excitation energy of the triplet spinon we
obtain by the Lanczos-Gagliano method agrees well with the Bethe-ansatz result
in general except that the energy of the triplet spinon II does not
seem to agree well when the
interaction is larger than $U/t=3$. From the result in Fig. 12(a),
we can see that this may be due to the appearance of another excitation
peak in the Lanczos-Gagliano spectra, which is not represented in 
our Bethe-ansatz
solutions. Based on these results we conclude that the triplet
spinon excitations are likely to be the dominant contributions
in the depolarized spectra.
In Fig. 13, we show the interaction dependence of the spectral weights of the
two triplet spinon excitations.
The triplet spinon I(II) has a maximum(minimum) spectral weight at
some finite interaction, $U/t=3\sim4$, and the interaction dependence of
the spectral weight is nontrivial. This result demonstrates the importance
of the intermediate interaction strength of
the 1D Hubbard model.
\subsection{Discussion}
We systematically study the elementary excitations of 1D
Hubbard model by combining the techniques of the exact Bethe-ansatz
equations for the mode dispersion and
the Lanczos-Gagliano method based spectral weight calculation
of the correlation functions. Three types of elementary
excitations, holon, singlet spinon, and triplet spinon excitations, are studied
at zero temperature and different densities ($\langle n\rangle$=5/6, 1/2 and
1/3) and different interaction strength, $U$. We first compare the
energy-momentum dispersion relations of these excitations obtained by both
methods and then study the mode
spectral weights in different situations.
The comparison between Bethe-ansatz solutions and resonance peaks of the
Lanczos-Gagliano correlation function gives us
some important results: (i) the holon and the singlet
spinon excitation states show up together in the
charge density correlation spectra. Holon states have higher energy with
an energy minimum at
$k=4k_F$ while the singlet spinons lie in
the lower energy region with an energy
minimum at $k=2k_F$. There are \textit{no} other states of
prominent spectral weights except the gapped double occupancy (soliton) states
near half filling. This result connects the
theoretical calculation of the 1D Hubbard
model with observable physical quantities ---
in particular, these are the only two modes which are likely to show up
in the polarized Raman scattering experiment probing the charge density
excitation spectra.
Another implication of this result is that one can interpret the
Bethe-ansatz
quantum numbers, $\{I_j\}$ and $\{J_\alpha\}$, as the ones of collective
excitations. But our spectral weight analysis shows that most of
the Bethe ansatz solutions for the 1D Hubbard chain
do not have any observable contributions to the
real physical quantities because they carry essentially no spectral weights.
(ii) The excitation holon II has a power-law behavior in its spectral weight
with respect to the interaction strength in the small $U/t$ region,
while the holon
I has almost interaction-independent spectral weight
(here the holon I/II could be generalized to
represent the $4k_F$-singlet excitations having holes in the edge/middle
of the charge quantum number, $\{I_j\}$). An interesting problem
for further research is to obtain an analytic formula for
the exponent of the holon II excitation. This will relate to the small
interaction expansion of Bethe ansatz equations and wave functions, which
have not yet been explored much in the literature.
When the interaction strength
increases, on the other hand, the spectral weights of
these two holons become equal as shown in Fig. 10(a). (iii)
As for the singlet spinons, we find that their spectral weights
decrease to zero very fast (exponentially) as $U$ increases. This
could be understood from the fact that the on-site repulsive interaction, $U$,
prevents the formation of the symmetric electron orbital wave functions,
which must accompany the antisymmetric spin singlet wave function, thus
suppressing the singlet spectral weight for large $U$.
(iv) From the imaginary part of the spin density correlation functions we
find that the triplet spinon is the only low energy spin
excitation in the long wavelength limit. There is \textit{no} other
excitation of important spectral weights in this region. (v) The spectral
weight study shows that among many 
triplet spinon excitation states, only those with some
some special quantum numbers could possibly have relatively
greater weights at
a given momentum (see Fig. 8(a) and the text related to that)
for finite interaction
strength, $U/t$. Others have very small or trivial weights, which are not
physically significant.
(vi) Finally
the interaction dependence of the spectral weights of the triplet spinon
I and II differs very much
in the intermediate interaction range, but becomes similar in magnitude
in both the weakly interacting
and the strongly interacting situations. This shows the subtle
complications in interpreting various excitation modes in the 1D Hubbard
model for intermediate interaction strength (say $U/t\sim 3$). Further
research is needed to provide a more complete understanding of this
intermediate interaction region, and our results should be considered
a preliminary investigation.
\section{CONCLUSION}
In this paper, we calculate the 1D charge and spin density correlation
functions (Eqs. (3) and (4)), which respectively correspond to the
polarized and the depolarized Raman scattering 
spectra, in three different models: 
Fermi liquid-RPA model, the Luttinger
liquid model, and the Hubbard model. Fermi liquid results (in RPA)
show a strong CDE mode (plasmon) with a very weak SPE mode in the polarized
spectra, while the Luttinger liquid results show the bosonic 
collective excitation 
(CDE or plasmon) as the only mode in the polarized spectra without 
any SPE. The 1D Hubbard model results show that both
holon and singlet spinon excitations contribute to the polarized spectra, and
we study their relative spectral weights and interaction dependence 
in detail. Note that the Hubbard model being a lattice model, its 
applicability to the RRS experiments in 1D semiconductor quantum wires (which
are continuum systems) is necessarily limited and is at best qualitative.

Focusing on the polarized RRS calculations, we can compare the 
lattice HM results with the continuum FL-RPA and LL results in the context
of experimental observations in 1D QWR systems~\cite{QWR_ref}.
According to the solutions of the HM Bethe-ansatz equations,
we find that the RRS spectrum is best interpreted as the combination of holon
and singlet spinon excitations as well as the CDE and "SPE" (or "SSE", i.e. the
spin singlet excitations)
in the high and low energy regions respectively.
Therefore, the numerical ratio of spectral weights of the singlet spinons
to the holons
should be interpreted as the ratio of the SSE(SPE) spectral weight with 
respect to the CDE spectral weight. The
experimental results show that SPE has almost the same spectral
weight as the CDE, i.e. this ratio is almost unity, whereas in our calculation,
this ratio is much smaller than unity
in the intermediate interaction range ($U/t>0.5$).
For weak interactions
($U/t<0.5$) it seems from Fig. 11 that SSE could have comparable weight to CDE. 
But this raises a serious problem with respect to 
the experimental results~\cite{QWR_ref}, which find  
the plasmon velocity to be at least two times greater than the Fermi 
velocity in the experimental long wavelength regime, while this is not
true for the 1D Hubbard chain unless $U/t>1$ as seen from Fig. 10(c).
Thus the plasmon dispersion (if we demand agreement between HM and the 
experimental results) indicates the appropriate interaction range
in the HM to be $U/t>1$, where our calculated 
SSE/CDE spectral weight ratio is extremely
small, in contrast to the experimental finding.
Therefore, we can conclude that 
the 1D Hubbard model results can not explain the experimental
data consistently for both the mode dispersion and
the spectral weight, although the lattice nature of the Hubbard model
makes a comparison with QWR experiments somewhat misleading.
Moreover we note that in the Hubbard model, the
on-site interaction,
$U$, is short-ranged (in fact, zero range), and operates only
between opposite spins at the same site.
We therefore expect that the spectral
weight of the singlet spinon is
(perhaps strongly) overestimated in the Hubbard model
compared with the Coulomb
interaction (which is explicitly spin-independent) case. We believe
the spectral weight of SSE (or SPE) in
a continuum interacting model should be even less
than that in the finite-sized Hubbard model here, just as we 
find in the RPA (or the HA) results
in the FL theory where the mode dispersion (both for CDE and SPE) agrees
well with experiments, but the spectral weight for the SPE is far too weak
in the theory. Based on this logic our
results seem to rule out the interpretation
that the SSE shows up as the additional anomalous
SPE mode in the RRS spectrum.
The singlet spin excitation according to our analysis could not provide
sufficiently large spectral weights in order to quantitatively explain
the large spectral weight in the so-called SPE mode seen in the
experimental RRS spectroscopy of 1D QWR systems.
This conclusion, as emphasized before, only applies to the NRS nature 
of our theory --- if the valence band and the resonant nature of 
RRS experiments is playing an important role, as likely
\cite{our_rrs_rpa,sassetti,our_rrs_ll}, then the theory needs 
to take resonance into account.

In summary, we systematically 
calculate the charge density (polarized spectra)
and the spin density correlation (depolarized spectra)
functions of one-dimensional systems in three different models:
Fermi liquid model, Luttinger liquid model, and Hubbard model.
In the polarized spectra, we find that the
FL model shows a strong collective charge density excitations at plasmon 
energy and a relatively weak single particle excitation at $\omega=qv_F$,
while the LL model shows one bosonic (plasmon/CDE) excitation only. 
Comparing the 
plasmon excitation energy of FL model and the bosonic excitation of LL model
(see Eqs. (11) and (17)) we find these two excitations are identical,
and the small SPE peak in FL model arises from the finite curvature
effect of electron energy dispersion at the Fermi point. 
In the Hubbard model, however, two excitations,
holons and singlet spinons, show up together in the polarized spectra.
We show that the holon excitations are actually the CDE in FL model or the 
bosonic excitation in LL model, while the singlet spinons in the HM arise from 
the spin degree of freedom and finite dispersion curvature at Fermi point.
If we compare the spectral weights of the lower energy excitations 
(SPE of FL model/no peak in LL model/singlet spinons in Hubbard model) 
and the weights of the higher energy 
excitations (CDE in FL model/boson peak in LL model/holons in Hubbard model), 
we find that the higher energy excitations always have (much) larger spectral 
weight than the lower energy ones in all models. This shows that the 
equal weight two-peak structure observed in the experiments~\cite{QWR_ref}
could not be explained by the \textit{nonresonant} 
Raman scattering mechanism, no
matter how one interprets the lower energy excitations to be 
SPE or SSE. Recent theoretical work~\cite{our_rrs_rpa,sassetti,our_rrs_ll}
on \textit{resonant} Raman scattering spectroscopy indicates that the low 
energy SPE feature may be a purely band structure effect arising from 
the participation of the valence band in the resonant scattering process. 
This also explains why this anomalous peak shows up in all dimensions
in experiments and not just in 1D.
In the depolarized spectra, however, only one spin excitation (the SDE
or the spin triplet excitation) is observed in 
these three models. The vertex correction of the FL model will 
in general reduce the 
SDE energy compared with the SPE energy, and separate the 
SDE from the SPE. In the intermediate
interaction region, the two triplet spinons in the Hubbard model have very 
different spectral weight behavior, showing very interesting interaction 
effects which need to be studied
in more details in the future. 

\textit{Acknowledgment}: The authors are very grateful to late Professor
Heinz Schulz for helpful correspondence on the one dimensional Hubbard model.
This work was supported by the US-ONR, and US-ARO.

\newpage
\begin{figure}

 \vbox to 5.2cm {\vss\hbox to 6cm
 {\hss\
   {\includegraphics{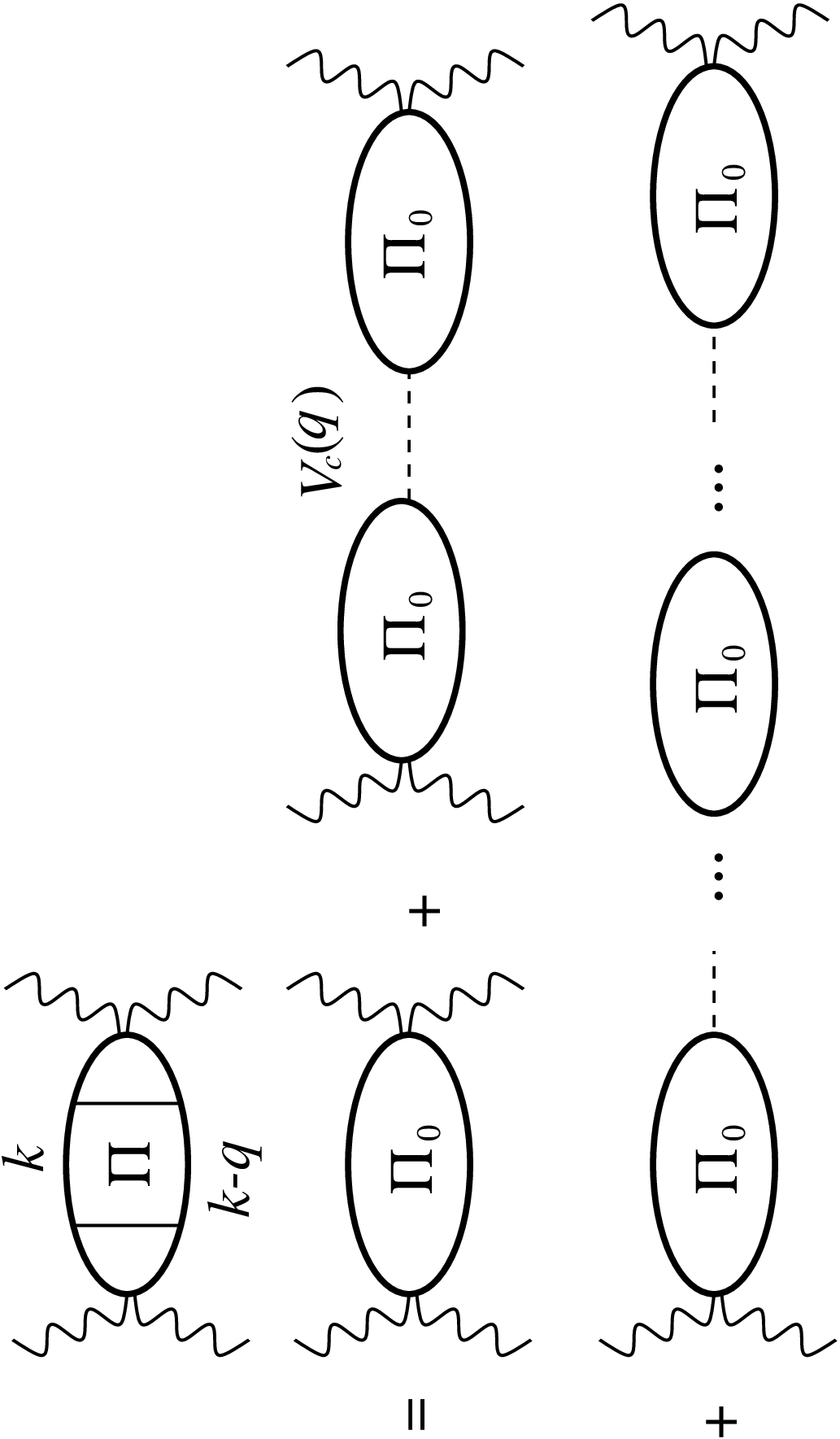}
   }
  \hss}
 }
\caption{
Diagrammatic representation of
the conduction band irreducible polarizability,
$\Pi_{0}(q,\omega)$ and reducible polarizability,
$\Pi(q,\omega)$, in standard RPA calculation.
$V_c(q)$ is the Coulomb interaction.
}
\end{figure}
\begin{figure}
 \vbox to 5.5cm {\vss\hbox to 6cm
 {\hss\
   {\includegraphics{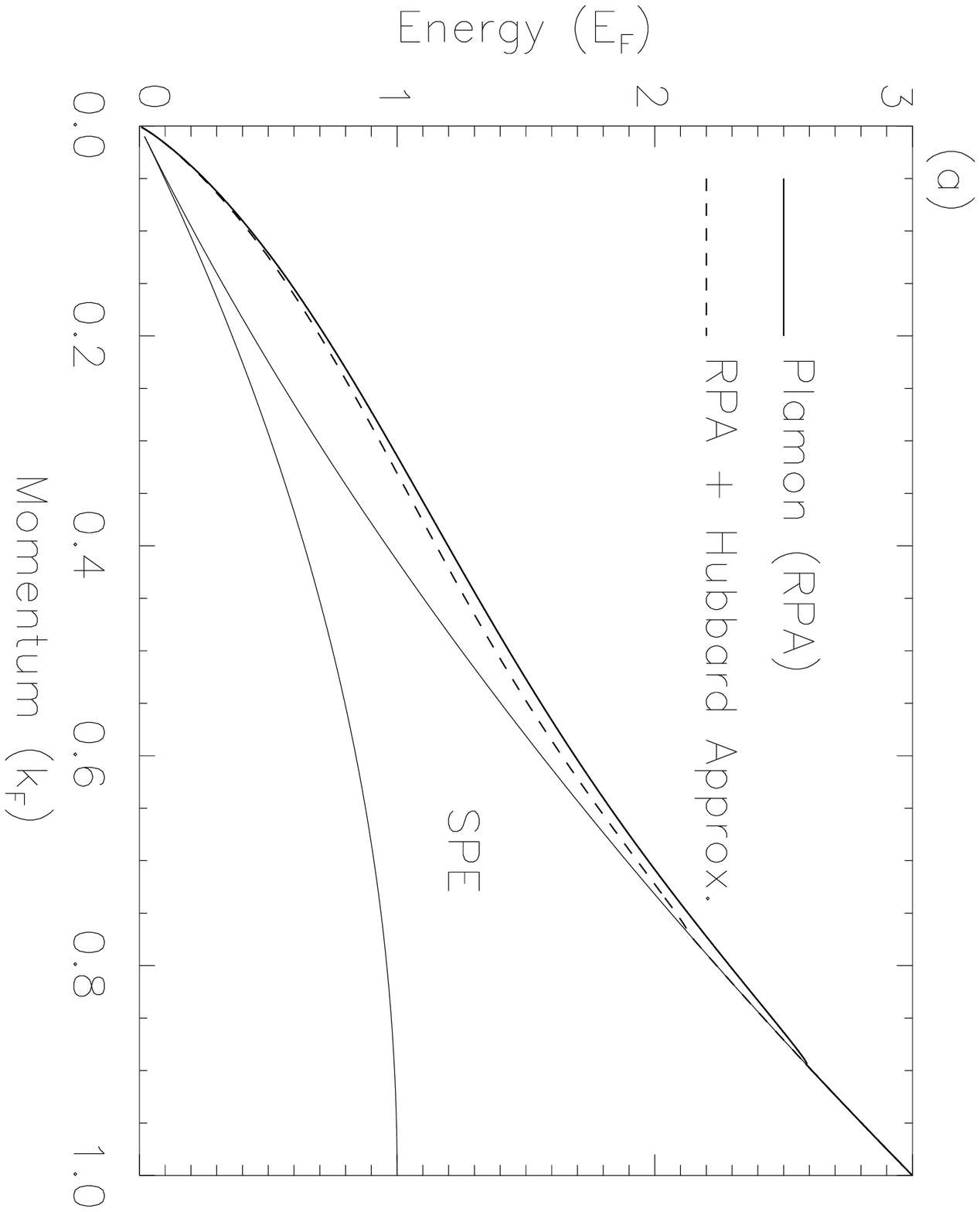}
   }
  \hss}
 }
 \vbox to 7cm {\vss\hbox to 10.cm
 {\hss\
   {\includegraphics{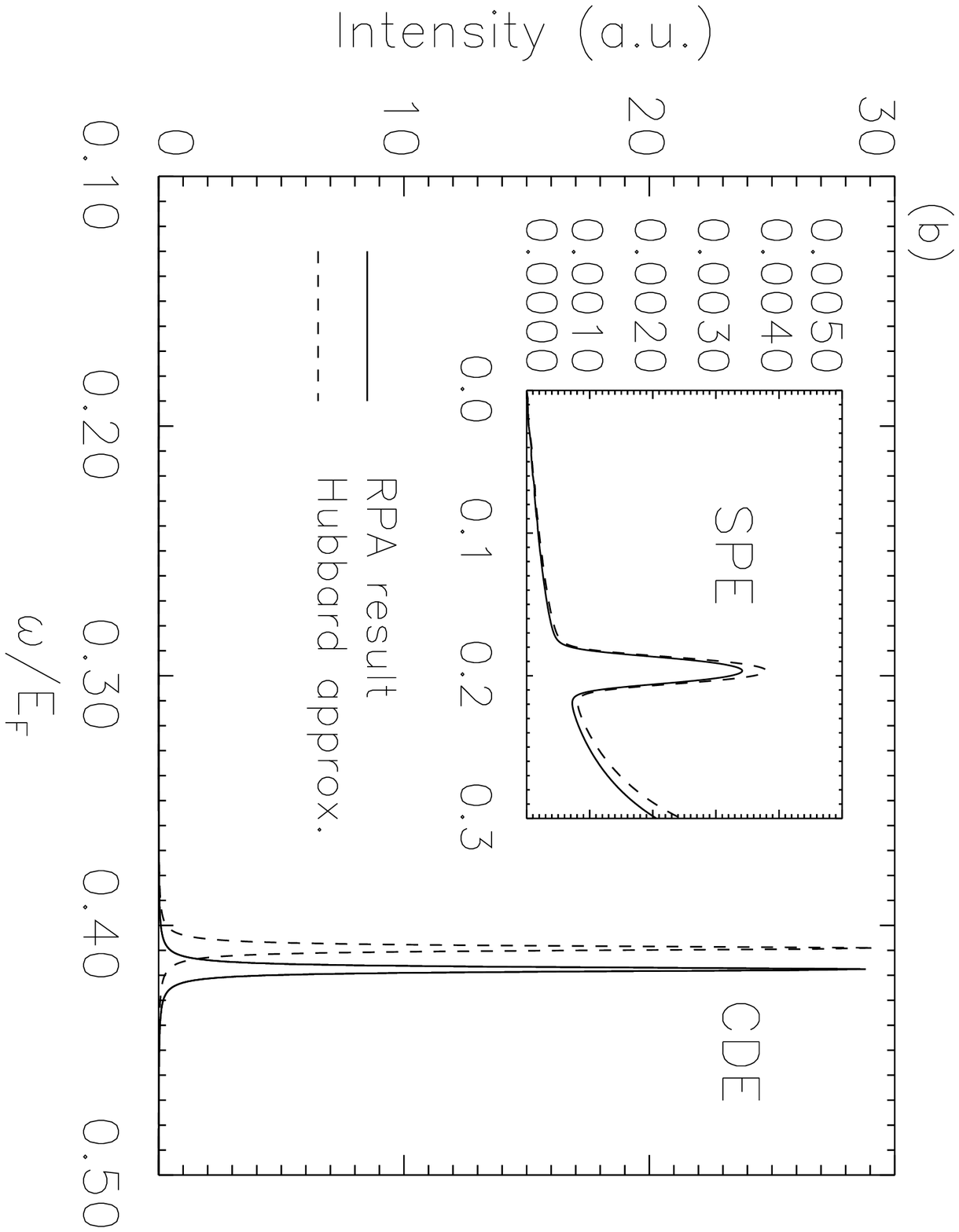}
   }
  \hss}
 }
\caption{
(a) The energy-momentum dispersion relation
for the plasmon mode and the SPE region of 1D system.
(b) The dynamical structure factor of the polarized RRS spectrum
in RPA calculation
for the 1D quantum wire system at $q=0.1k_F$. Vertex correction in
Hubbard approximation is also shown for comparison.
Parameters are the same as the experiments in ref. [4].
Finite broadening factor is involved to present the delta-function peaks.
}
\end{figure}
\begin{figure}
 \vbox to 6cm {\vss\hbox to 6cm
 {\hss\
   {\includegraphics{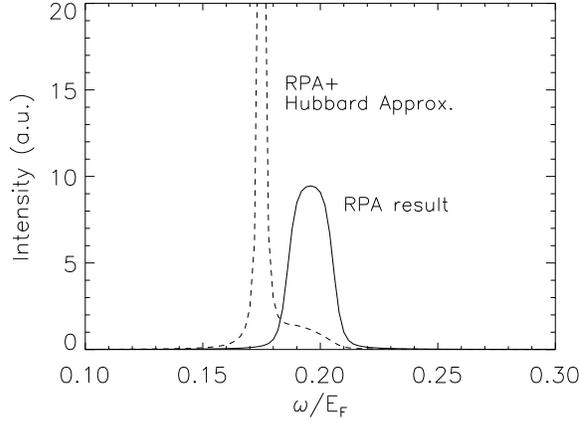}
   }
  \hss}
 }
\caption{
The dynamical structure factor of the depolarized RRS spectrum calculated
from Im$\Pi_{0,\sigma}^{\rm{RPA}}(q,\omega)$ 
within RPA and Hubbard approximation.
}
\end{figure}
\begin{figure}
 \vbox to 5.5cm {\vss\hbox to 6cm
 {\hss\
   {\includegraphics{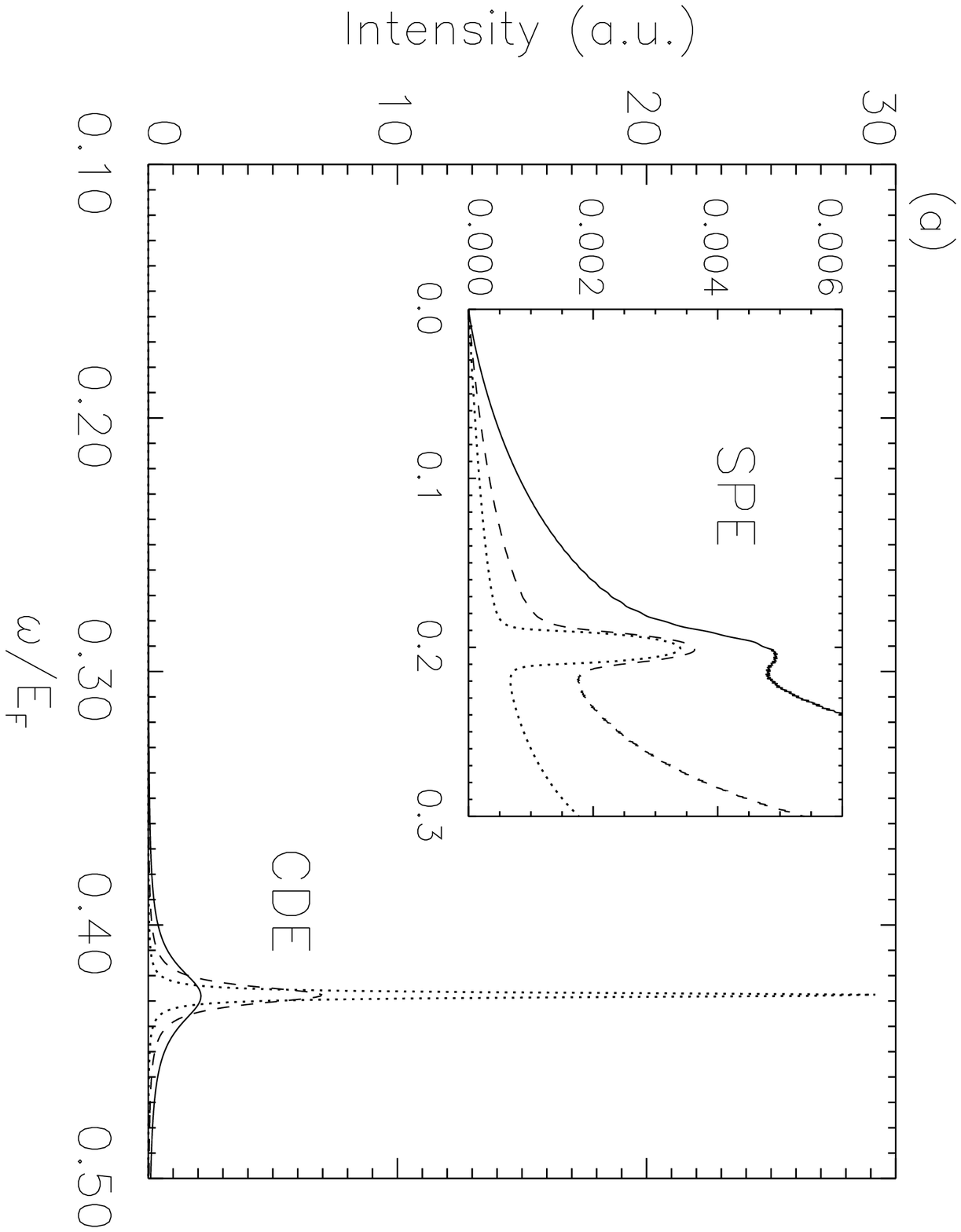}
   }
  \hss}
 }
 \vbox to 7cm {\vss\hbox to 10.cm
 {\hss\
   {\includegraphics{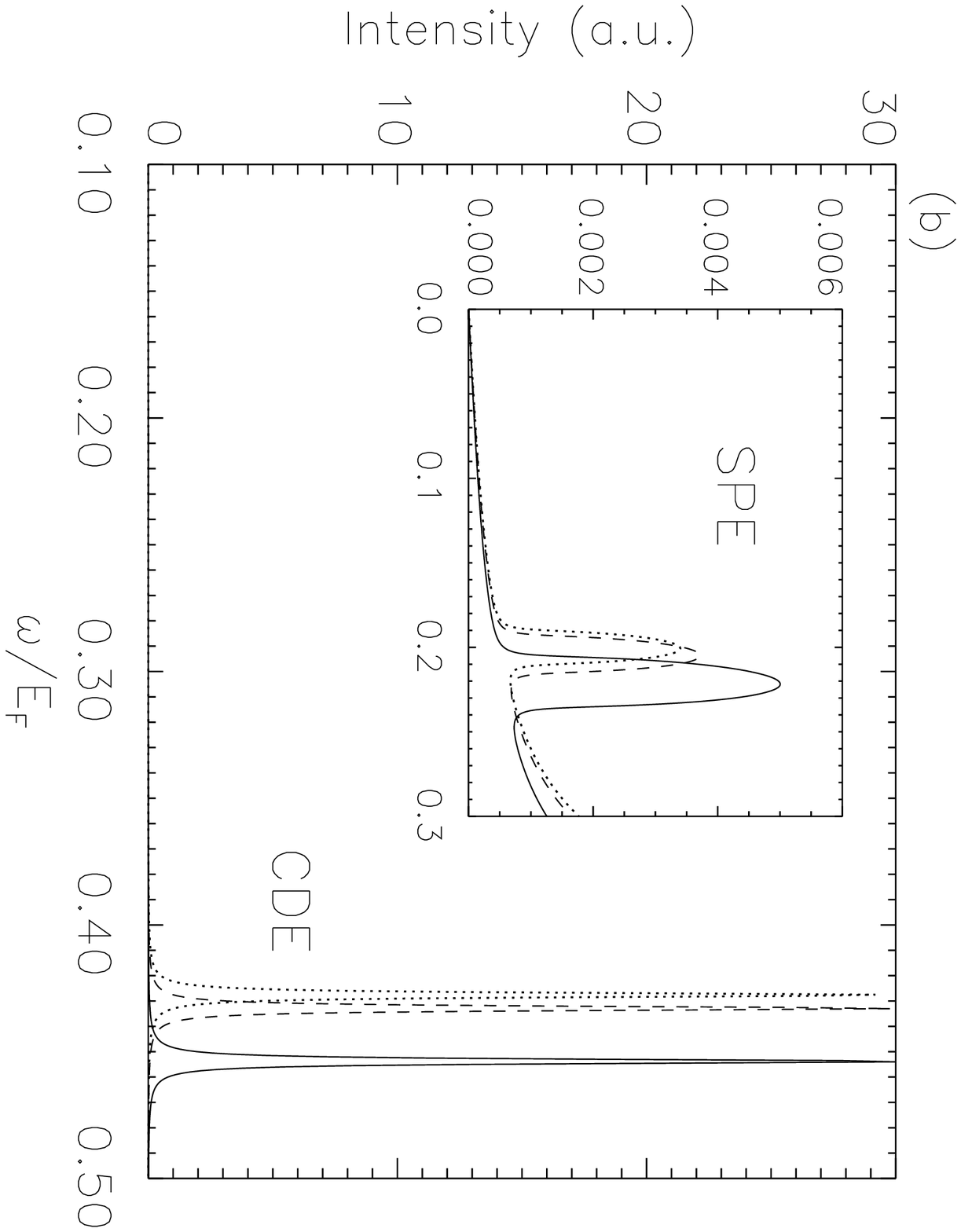}
   }
  \hss}
 }
\caption{
The dynamical structure factor of the polarized RRS spectrum
calculated by including (a) the breakdown of momentum conservation
and (b) the nonparabolic energy dispersion. The dot, dashed, and solid
lines in (a) represent the broadening parameter $\Gamma=0$, $10^{-3}$, and
$4\times10^{-3}$ respectively (see Eq. (12)),
and in (b) represent the nonparabolicity parameter
$\lambda=0$, 0.02, and 0.1 respectively (see Eq. (13)). All these effects
cannot enhance the SPE spectral weight to be comparable to CDE
in the reasonable range of $\Gamma$ or $\lambda$.
}
\end{figure}
\begin{figure}

 \vbox to 6cm {\vss\hbox to 6cm
 {\hss\
   {\includegraphics{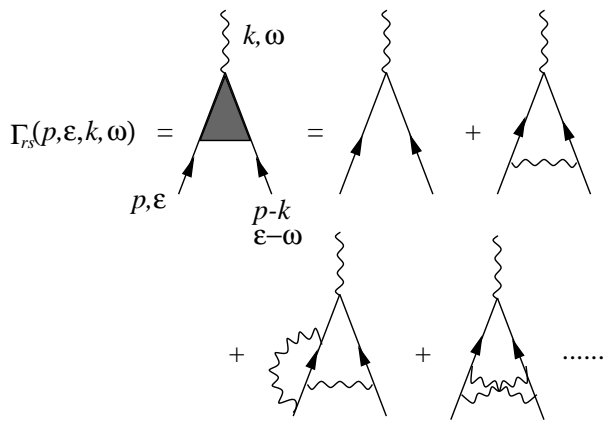}
   }
  \hss}
 }
\caption{
Diagrammatic representation of the Ward identity for
the vertex function $\Gamma_{rs}$. The solid lines represent the
single-particle Green's function while the wave lines represent the
interaction.
}
\end{figure}
\begin{figure}

 \vbox to 8cm {\vss\hbox to 6cm
 {\hss\
   {\includegraphics{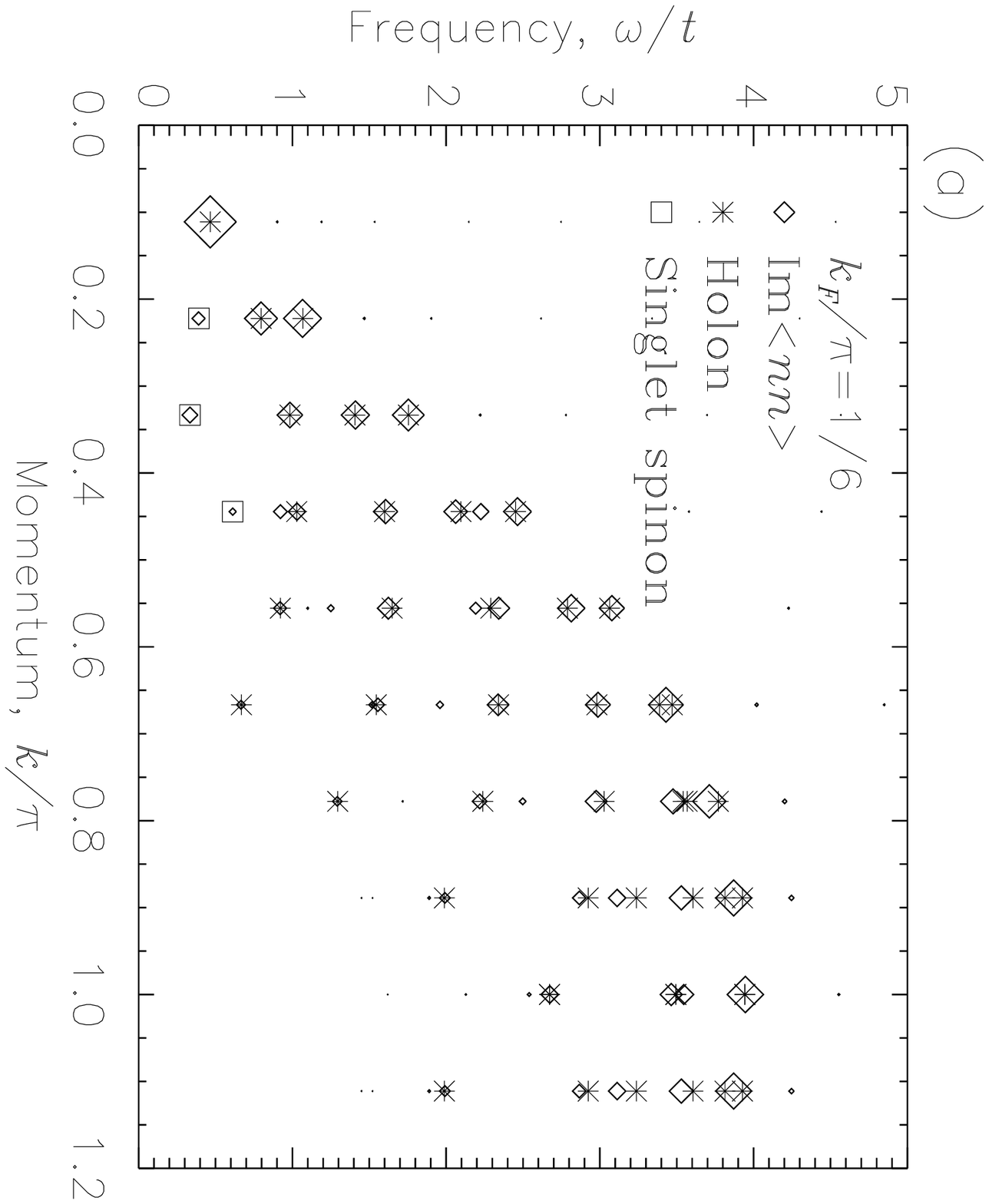}
   }
  \hss}
 }
 \vbox to 5cm {\vss\hbox to 6cm
 {\hss\
   {\includegraphics{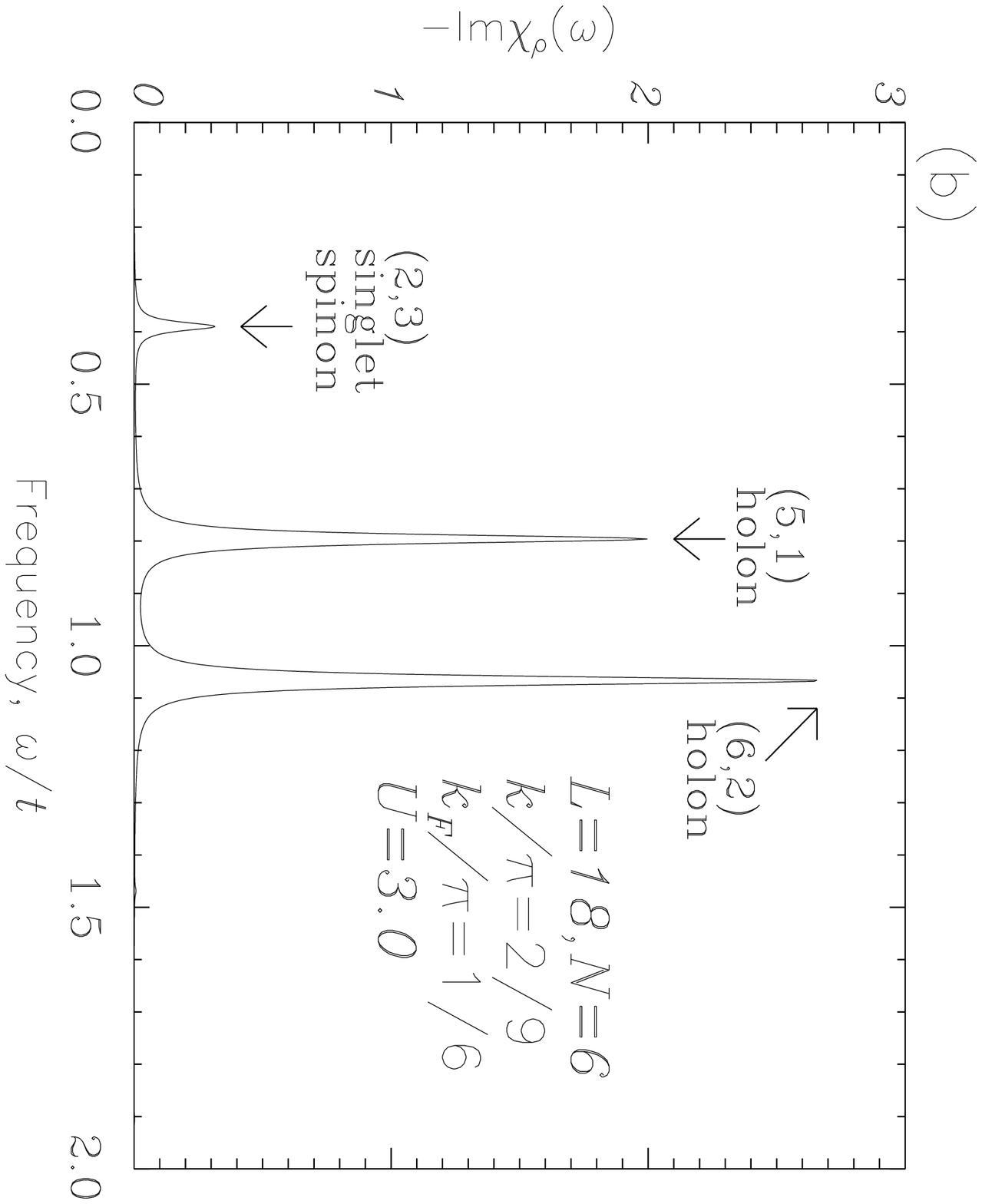}
   }
  \hss}
 }
\caption{
(a) Energy-momentum dispersion relation and (b) the
spectrum of charge density correlation function
for 6 electrons in 18 sites.
The area of each diamond(square) in (a) is proportional to the
spectral weight of each charge(spin) excitation peak. The numbers above
the holon
and singlet spinon peaks in (b) are the quantum numbers defined in
Sec. IV-A1 from the Bethe-ansatz equations.
}
\end{figure}
\begin{figure}

 \vbox to 8cm {\vss\hbox to 6cm
 {\hss\
   {\includegraphics{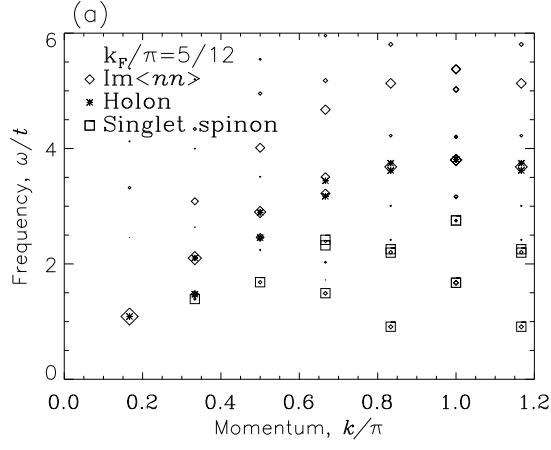}
   }
  \hss}
 }
 \vbox to 5cm {\vss\hbox to 6cm
 {\hss\
   {\includegraphics{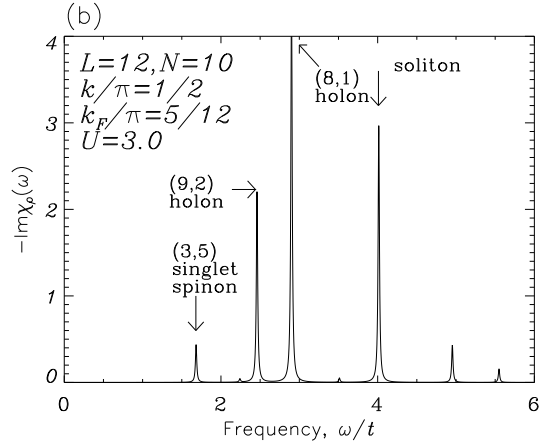}
   }
  \hss}
 }
\caption{
Same as Fig. 6, but for 10 electrons in 12 sites. One can see that the double
occupancy of electrons will give higher energy excitations in this high density
system. Even the singlet spinon excitations span in a larger area in the whole
momentum range in (a), their spectral weights
are still smaller than the holons and solitons (double occupancy excitations).
}
\end{figure}
\begin{figure}

 \vbox to 8cm {\vss\hbox to 6cm
 {\hss\
   {\includegraphics{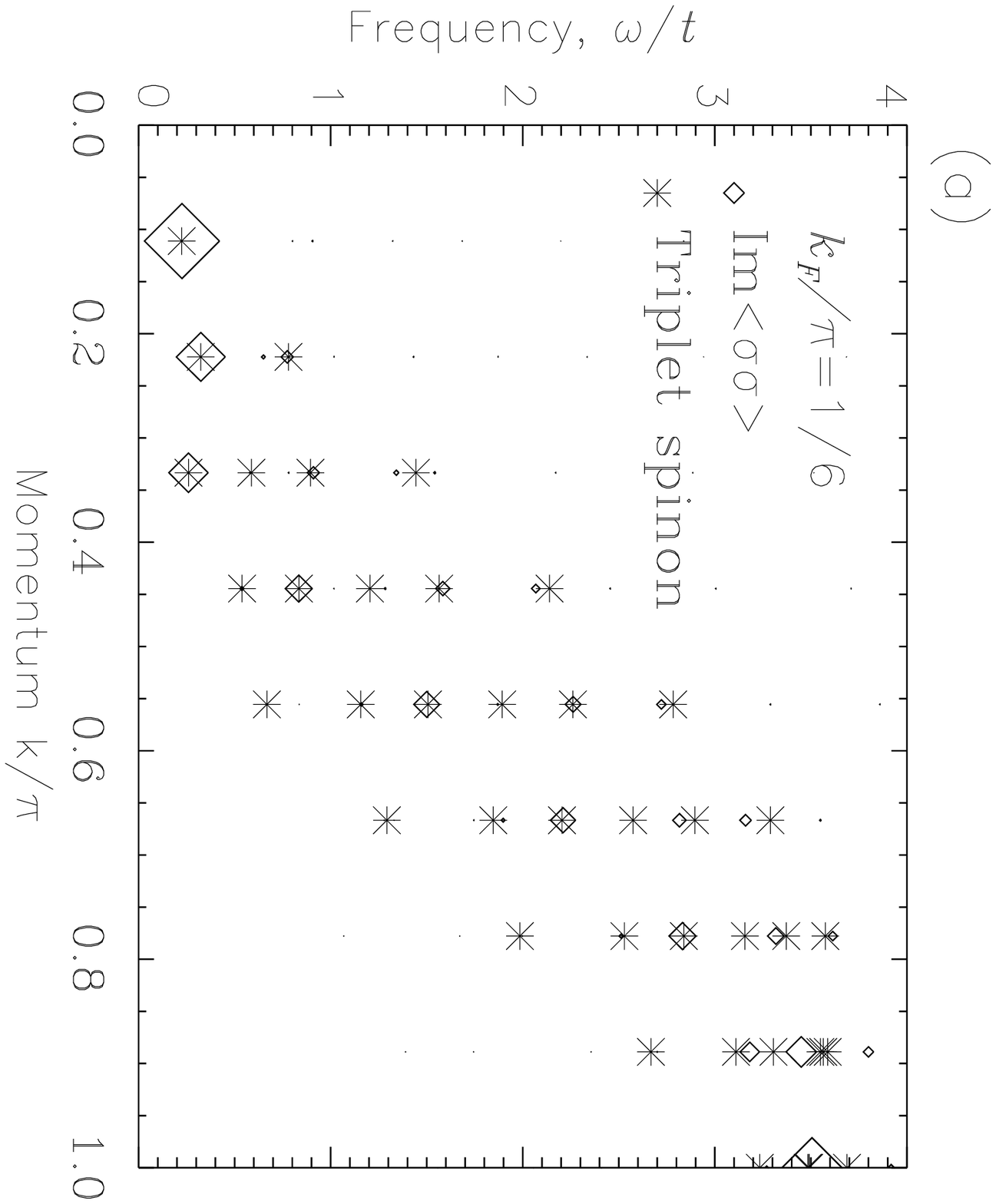}
   }
  \hss}
 }
 \vbox to 4.cm {\vss\hbox to 6cm
 {\hss\
   {\includegraphics{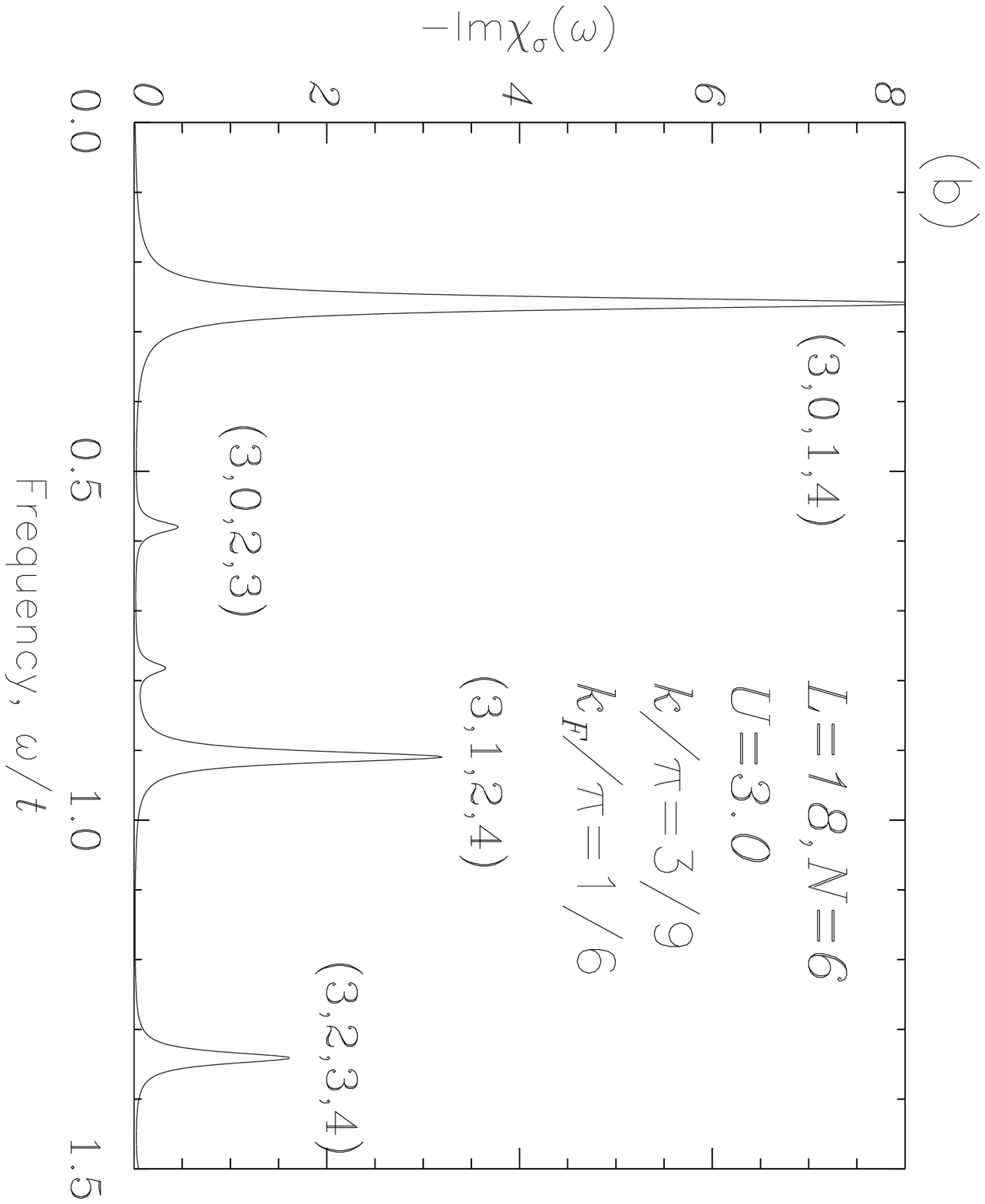}
   }
  \hss}
 }
\caption{
(a) Energy-momentum dispersion relation and (b) the
spectrum of the spin correlation function for
6 electrons in 18 sites.
}
\end{figure}
\begin{figure}

 \vbox to 6cm {\vss\hbox to 6cm
 {\hss\
   {\includegraphics{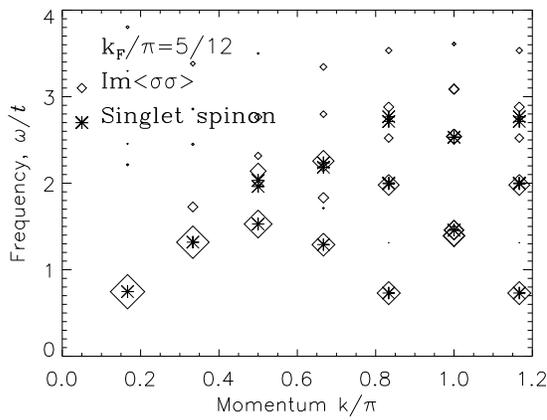}
   }
  \hss}
 }
\caption{
Same as Fig. 8(a), but for 10 electrons in 12 sites.
}
\end{figure}
\begin{figure}

 \vbox to 12cm {\vss\hbox to 6cm
 {\hss\
   {\includegraphics{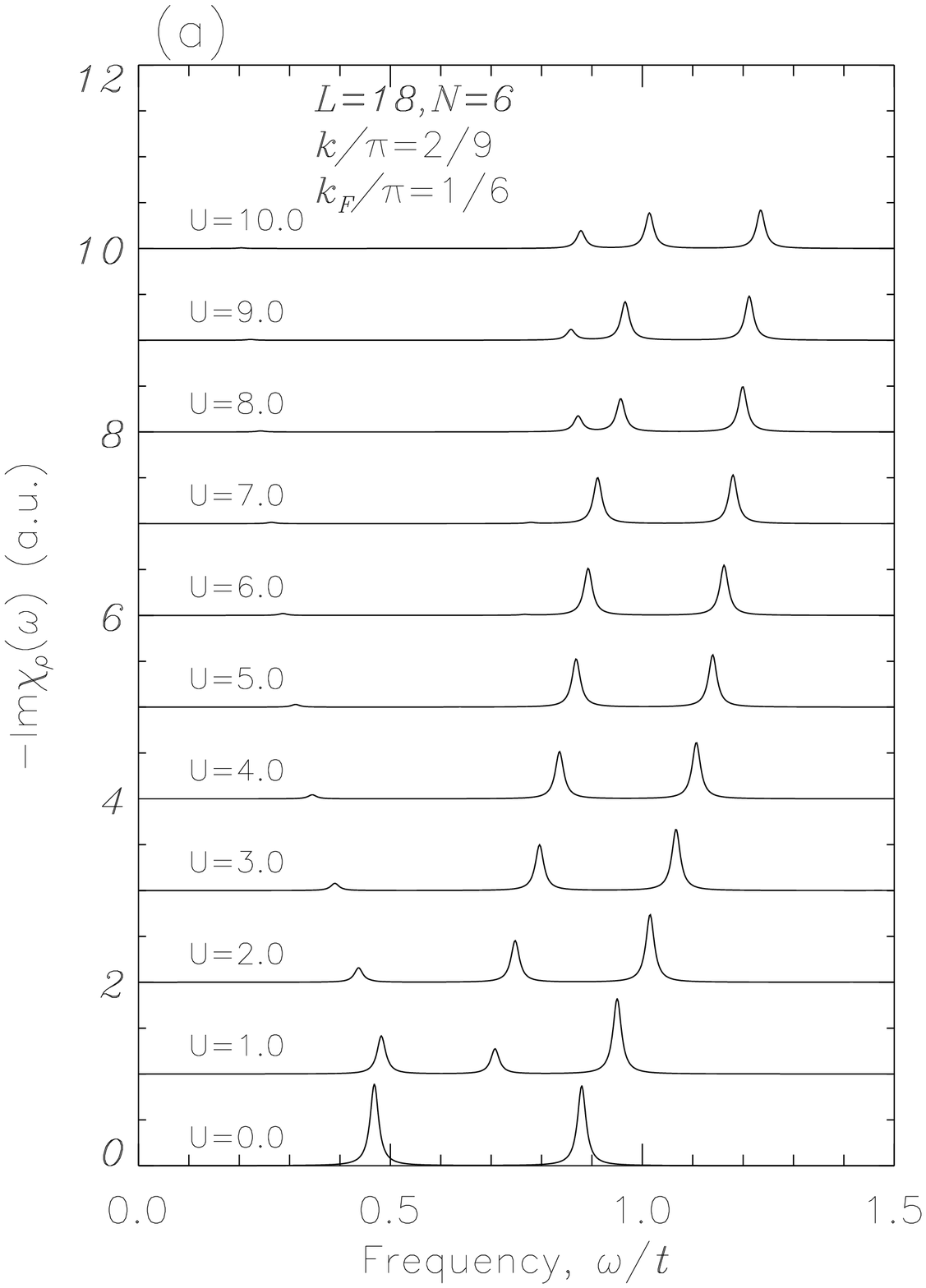}
   }
  \hss}
 }
 \vbox to 2.5cm {\vss\hbox to 6cm
 {\hss\
   {\includegraphics{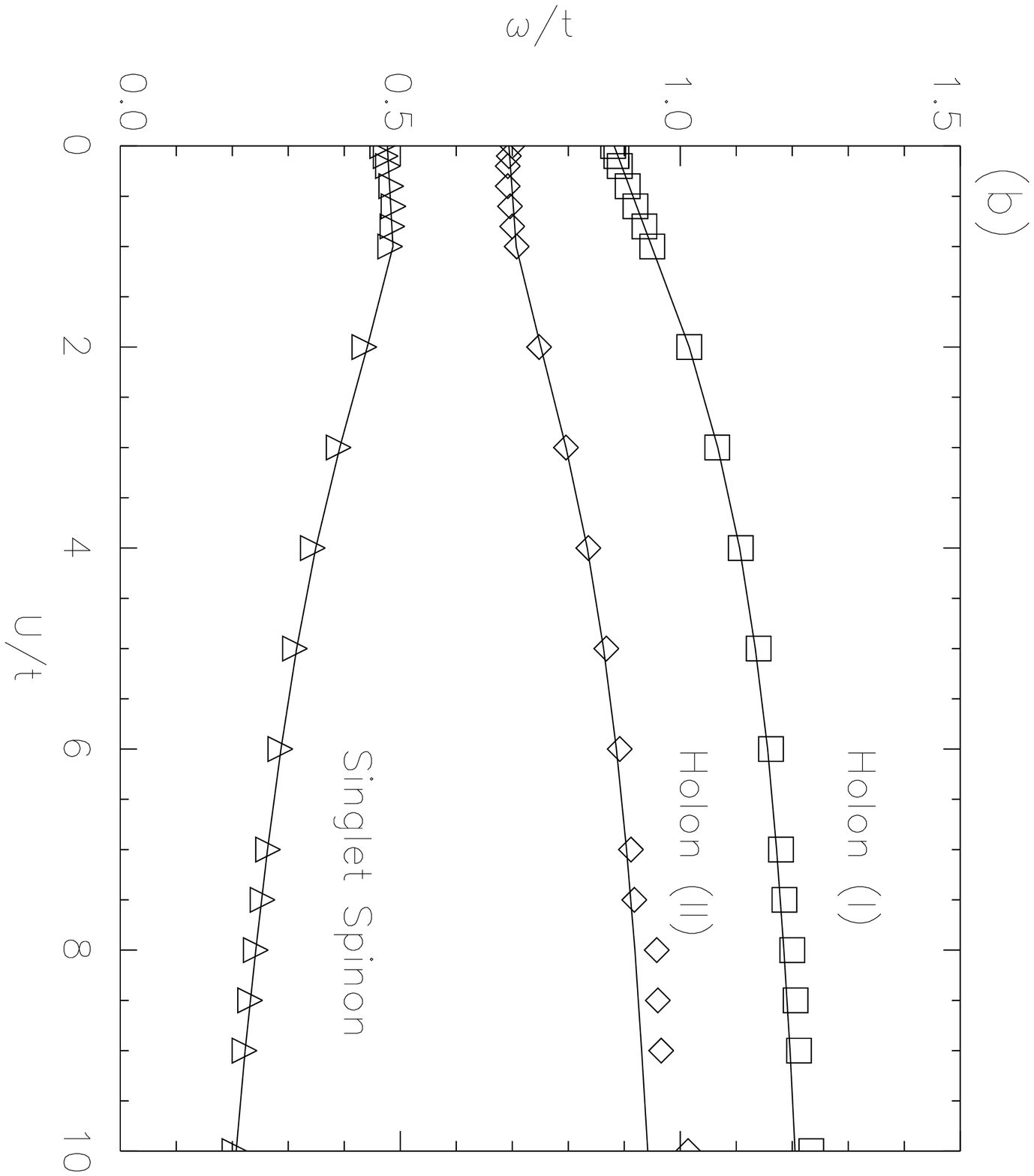}
   }
  \hss}
 }
 \vbox to 4.cm {\vss\hbox to 6cm
 {\hss\
   {\includegraphics{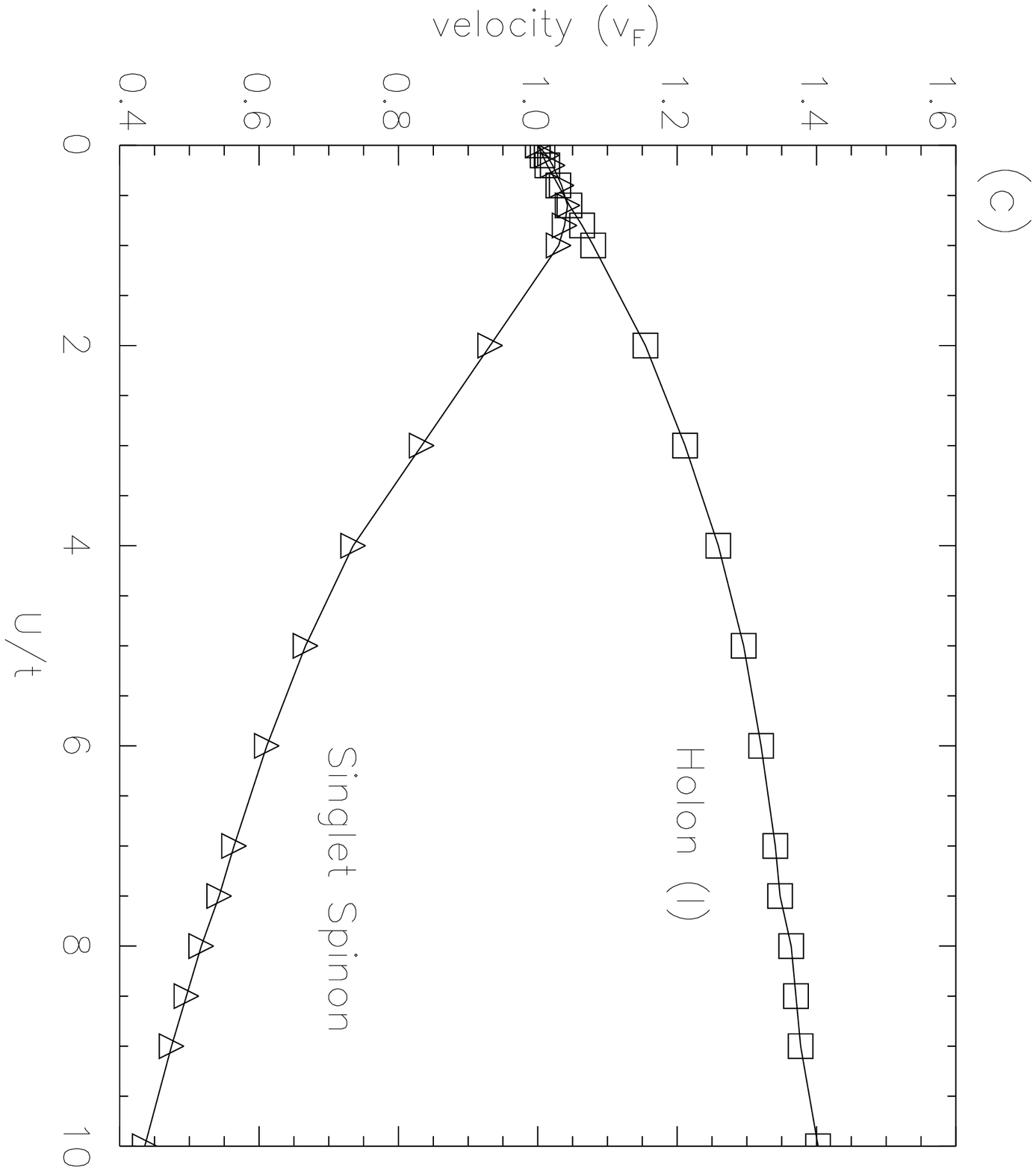}
   }
  \hss}
 }
\caption{(a) The calculated polarized spectra with various interaction
strengths for $\langle n\rangle=6/18=1/3$, at
$k=2\pi/9$, and (b) the excitation energies of the three elementary excitations
in various $U/t$, compared with the Bethe-ansatz results
(solid lines).
(c) The velocities of holon and singlet spinon
with respect to the Fermi velocity, $v_F$,
as a function of $U/t$.
}
\end{figure}
\begin{figure}

 \vbox to 8cm {\vss\hbox to 6cm
 {\hss\
   {\includegraphics{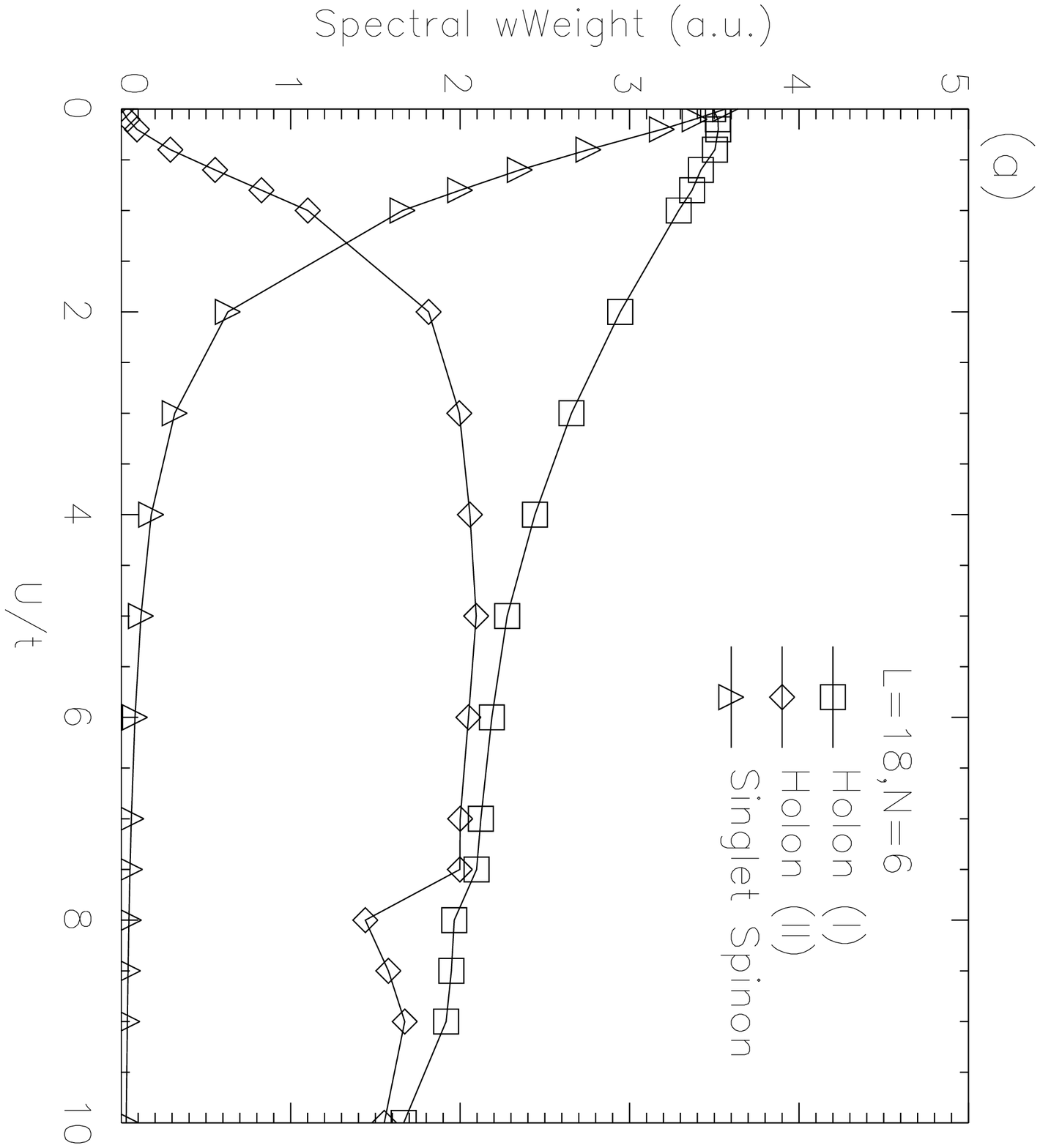}
   }
  \hss}
 }
 \vbox to 4cm {\vss\hbox to 6cm
 {\hss\
   {\includegraphics{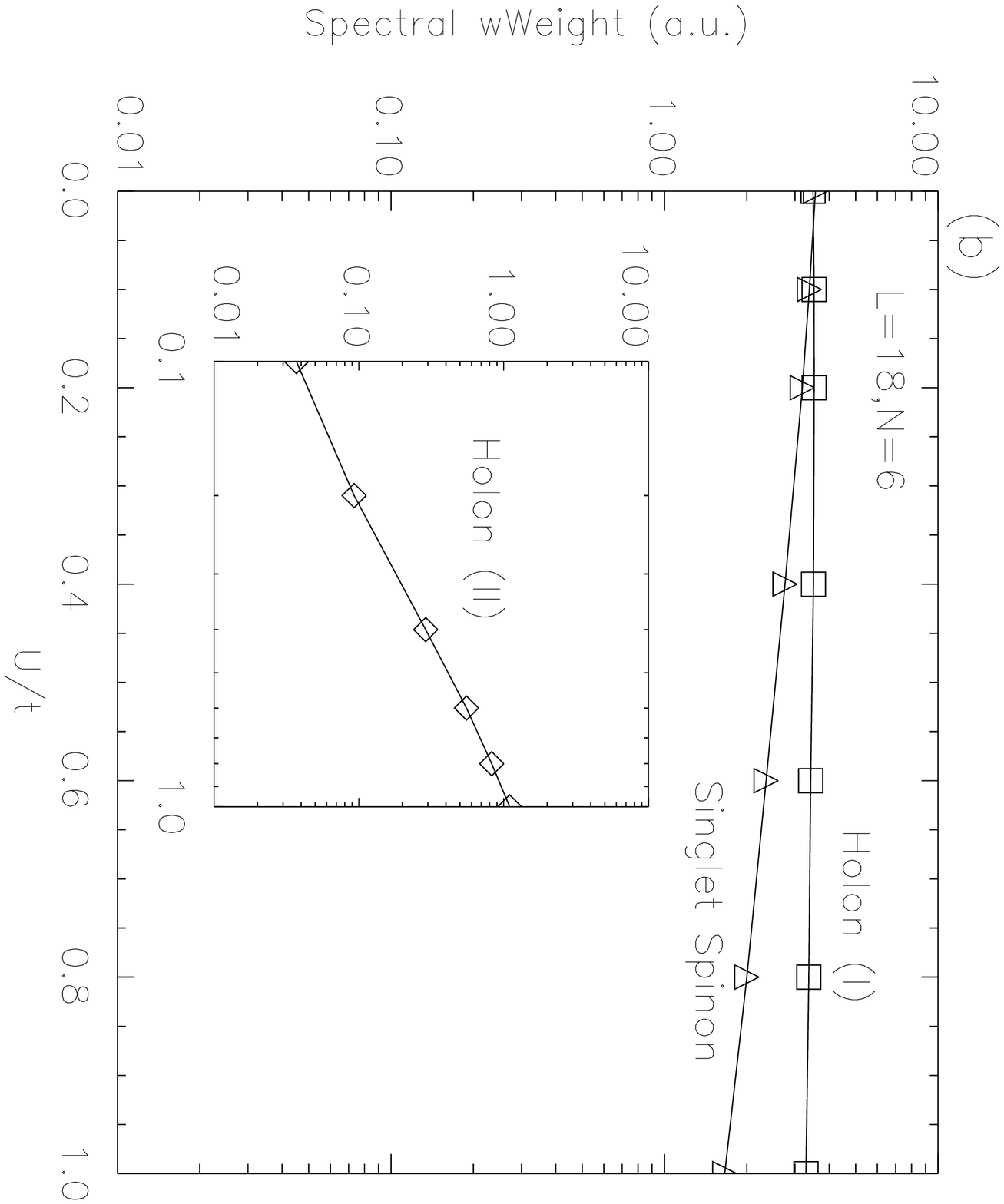}
   }
  \hss}
 }
\caption{The spectral weights of the three excitations in the polarized spectrum
as a function of the interaction strength, $U/t$, in (a) linear scale
from $U/t=0$ to $U/t=10$, and (b) linear-log scale for
small $U$ ($U/t\le 1.0$). The momentum $k=2\pi/9$ is the same as in Fig.
6. The inset of (b) is the log-log plot for the holon II excitation.
We can see that the holon II excitation increases power-lawly in its strength
as $U$ increases, showing a possible Luttinger liquid behavior in
the weakly interacting system (see text).
}
\end{figure}
\begin{figure}

 \vbox to 14cm {\vss\hbox to 6cm
 {\hss\
   {\includegraphics{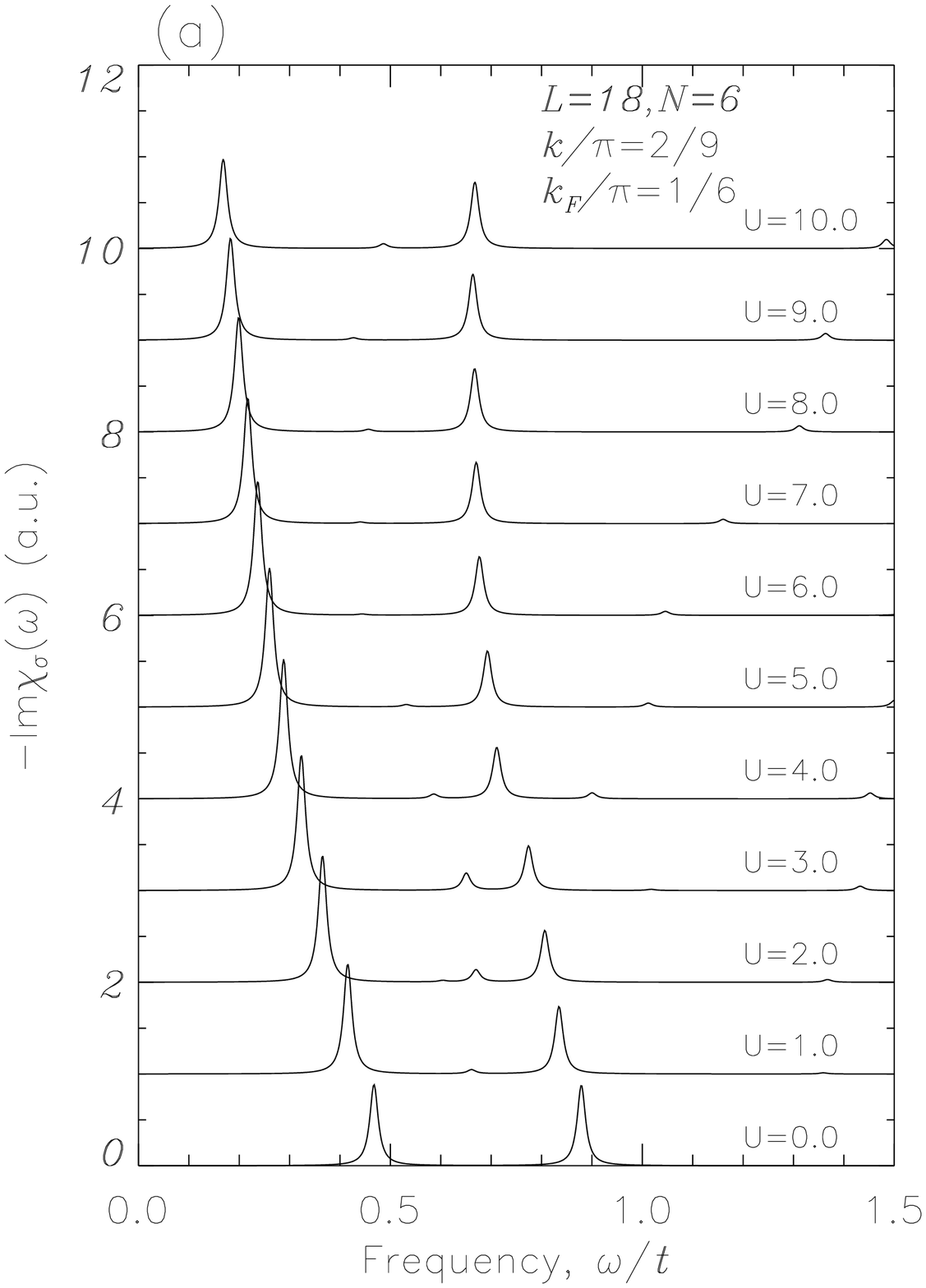}
   }
  \hss}
 }
 \vbox to 1cm {\vss\hbox to 6cm
 {\hss\
   {\includegraphics{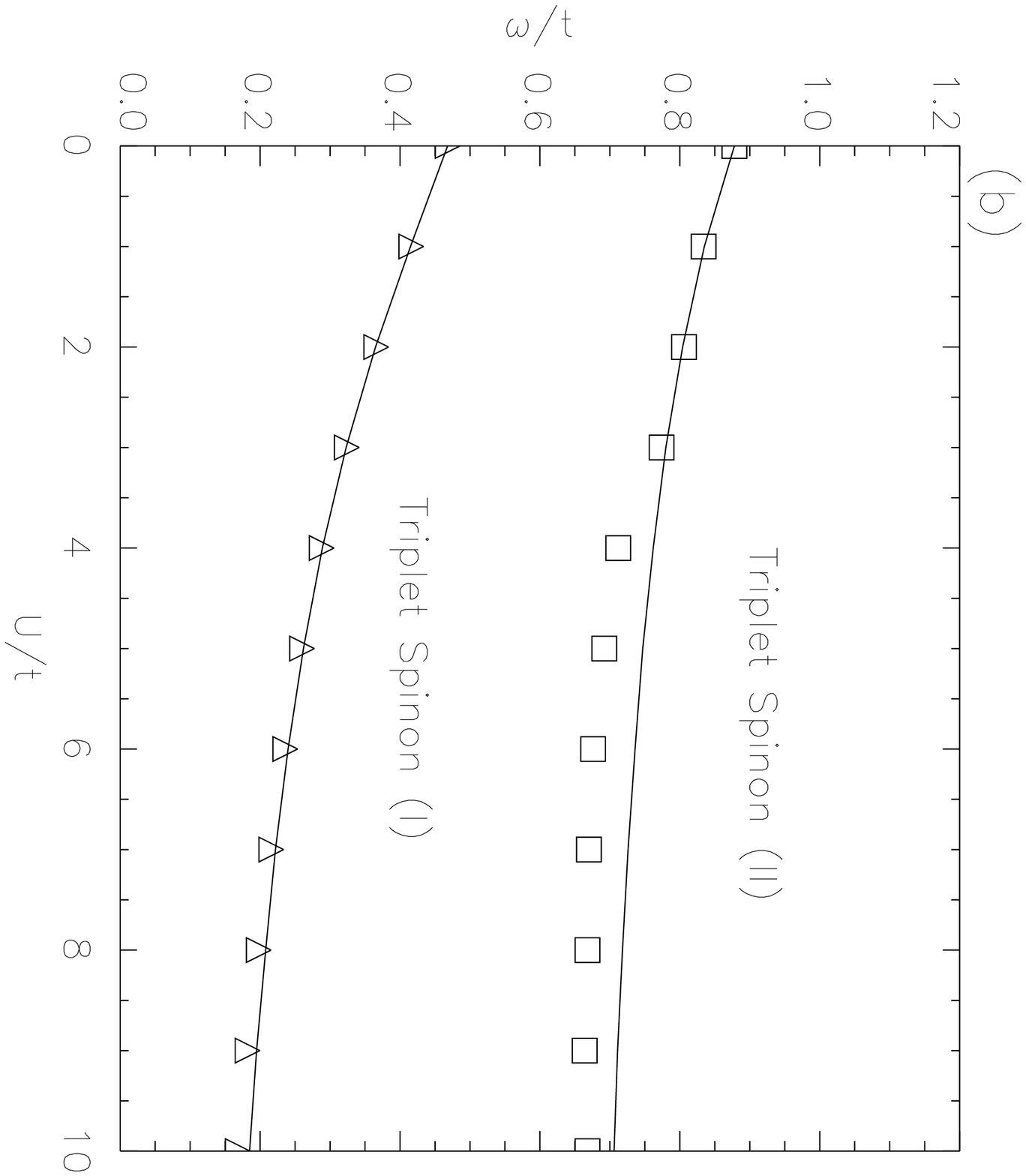}
   }
  \hss}
 }
\caption{(a) The calculated depolarized spectra with various interaction
strengths of the low density system,
$\langle n\rangle=6/18=1/3$, at momentum
$k=2\pi/9$, and (b) the excitation energies of the two elementary excitations
(triplet spinon I and II) with respect to the interaction strength, $U/t$,
compared with the Bethe-ansatz result (solid lines).
}
\end{figure}
\begin{figure}

 \vbox to 5.8cm {\vss\hbox to 6cm
 {\hss\
   {\includegraphics{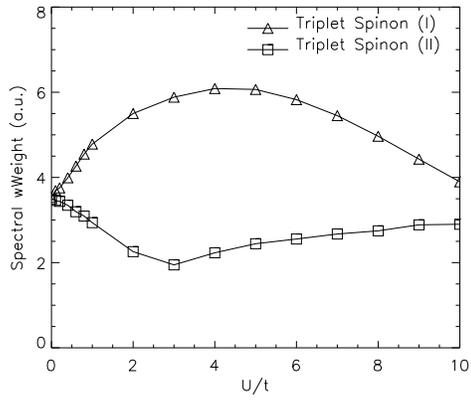}
   }
  \hss}
 }
\caption{The spectral weights of the triplet spinon
excitations as a function of interaction,
$U/t$, in linear scale from $U/t=0$ to $U/t=10$.
We consider the same system and momentum as in Fig. 12.
}
\end{figure}
\end{document}